\documentclass[nofootinbib,aps,pra,twocolumn,showpacs,groupaddress,preprintnumbers,floatfix]{revtex4-1}

\usepackage{dynlearn}

\begin{document}

\def\ourTitle{
Strong and Weak Optimizations\\
in Classical and Quantum Models\\
of Stochastic Processes
}

\def\ourAbstract{
Among the predictive hidden Markov models that describe a given stochastic
process, the \eM\ is strongly minimal in that it minimizes every R\'enyi-based
memory measure. Quantum models can be smaller still. In contrast with the \eM's
unique role in the classical setting, however, among the class of processes
described by pure-state hidden quantum Markov models, there are those for which
there does not exist any strongly minimal model. Quantum memory optimization
then depends on which memory measure best matches a given problem circumstance.
}

\def\ourKeywords{
  stochastic process, hidden Markov model, \texorpdfstring{\eM}{epsilon-machine}, causal states, quantum information.
}

\hypersetup{
  pdfauthor={James P. Crutchfield},
  pdftitle={\ourTitle},
  pdfsubject={\ourAbstract},
  pdfkeywords={\ourKeywords},
  pdfproducer={},
  pdfcreator={}
}

\author{Samuel Loomis}
\email{sloomis@ucdavis.edu}

\author{James P. Crutchfield}
\email{chaos@ucdavis.edu}
\affiliation{Complexity Sciences Center and Physics Department,
University of California at Davis, One Shields Avenue, Davis, CA 95616}

\date{\today}
\bibliographystyle{unsrt}

\title{\ourTitle}

\begin{abstract}
\ourAbstract
\end{abstract}



\preprint{\arxiv{1809.XXXX}}

\title{\ourTitle}
\date{\today}
\maketitle

\setstretch{1.1}

\listoffixmes

\section{Introduction}
\label{sec:introduction}

When studying classical stochastic processes, we often seek models and
representations of the underlying system that allow us to simulate and predict
future dynamics. If the process is memoryful, then models that generate it or
predict its future actions must also have memory. Memory, however, comes at
some resource cost; both in a practical sense---consider, for instance, the
substantial resources required to generate predictions of weather and climate
\cite{Lore63a,Lore64a}---and in a theoretical sense---seen in analyzing
thermodynamic systems such as information engines \cite{Boyd15a}. It is
therefore beneficial to seek out a process' minimally resource-intensive
implementations.

Predicting and simulating classical processes, and monitoring the memory
required, led to a generalization of statistical mechanics called
\emph{computational mechanics} \cite{Crut88a,Crut92c,Shal98a,Crut12a}. To date
computational mechanics focused on discrete stochastic processes. These are
probability measures $\mathbb{P}\left(\dots x_{-1} x_0 x_1 \dots\right)$ over
strings of symbols taking values in a finite alphabet $\mathcal{A}$. The
minimal information processing required to predict the sequence is represented
by a type of hidden Markov model called the \emph{\eM}. The statistical
complexity $\Cmu$---the memory rate for \eMs\ to simultaneously generate many
copies of a process---is a key quantity and a proposed invariant for measuring
the process' structural complexity.

When simulating classical processes, quantum systems can be constructed that
have smaller memory requirements than the \eM\ \cite{Gu12a,Maho15a}. The
$q$-machine is a particular implementation of quantum simulation that has shown
advantage in memory rate over a wide range of processes; often the advantage is
unbounded \cite{Agha17a,Suen17a,Garn17a,Agha18a}. For quantum models, the
minimal memory rate $C_q$ has been determined in cases such as the Ising model
\cite{Suen17a} and the Perturbed Coin Process \cite{Thom18a}, where the
$q$-machine attains the minimum rate. And so, though a given $q$-machine's
$C_q$ can be readily calculated \cite{Riec15b}, in many cases the absolute
minimum $C_q$ is not known.

Properly accounting for memory requires an appropriate formalism for resources
themselves. The field of resource theory has recently emerged in quantum
information theory as a toolkit for addressing resource consumption in the
contexts of entanglement, thermodynamics, and numerous other quantum and
classical resources \cite{Coec16a}. Its fundamental challenge is to determine
when one system, or resource, can be converted to another using a predetermined
set of \emph{free} operations.

\begin{figure}
\centering
\includegraphics[width=\columnwidth]{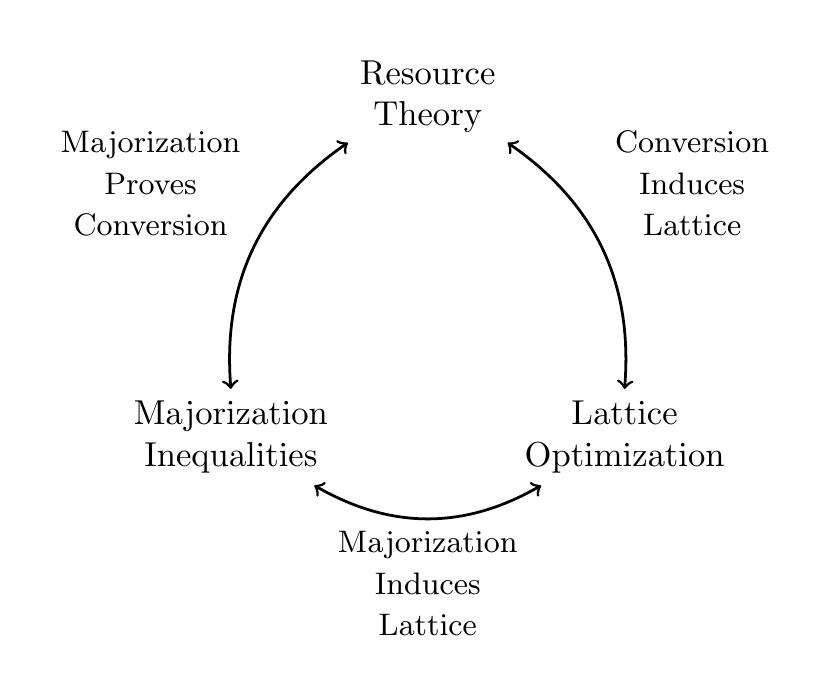}
\caption{Triumvirate of resource theory, majorization, and lattice theory.
	}
\label{fig:triad}
\end{figure}

Resource theory is closely allied with two other areas of mathematics, namely
majorization and lattice theory. Figure \ref{fig:triad} depicts their
relationships.

On the one hand, majorization is a preorder relation $\succsim$ on positive
vectors (typically probability distributions) computed by evaluating a set of
inequalities \cite{Mars11a}. If the majorization relations hold between two
vectors, then one can be converted to the other using a certain class of
operations. Majorization is used in some resource theories to numerically test
for convertibility between two resources \cite{Niel99a,Horo13a,Gour15a}.

Lattice theory, on the other hand, concerns partially ordered sets and their suprema
and infima, if they exist \cite{Grat10a}. Functions that quantify the practical
uses of a resource are monotonic with respect to the partial orders induced by
convertibility and majorization. Optimization of practical measures of memory
is then related to the problem of finding the extrema of the lattice.
Majorization and resource convertibility are both relations that generate
lattice-like structures on the set of systems.

Here, we examine the memory costs of classical and quantum models of stochastic
processes via majorization. Using lattice-theoretic intuition, we then define
the concept of \emph{strong optimization}, which occurs when a particular model
simultaneously optimizes all measures of memory via its extremal position in
the lattice. We show that among classical predictive models, the \eM is
strongly minimal. Following this, we show that the \eM is strongly maximal to a
subset of quantum models but that no strongly minimal quantum model exists in
some circumstances. These results constitute initial steps to a resource theory
of memoryful information processing.

\section{Majorization and optimization}
\label{sec:Majorization}

The majorization of positive vectors provides a qualitative description of how
concentrated the quantity of a vector is over its components. For ease of
comparison, consider vectors $\mathbf{p}=(p_i)$, $i\in\{1,\dots,n\}$, whose
components all sum to some constant value, which we take to be unity:
\begin{align*}
\sum_{i=1}^n p_i = 1
  ~,
\end{align*}
and are nonnegative: $p_i \geq 0$. For our purposes, we interpret these vectors
as probability distributions.

Our introduction to majorization here follows Ref. \cite{Mars11a}. The
historical definition of majorization is also the most intuitive, starting with
the concept of a \emph{transfer operation}.

\begin{Def}[Transfer operation]
A \emph{transfer operation} $\mathbf{T}$ on a vector $\mathbf{p}=(p_i)$
selects two indices $i,j\in\{1,\dots,n\}$, such that $p_i>p_j$,
and transforms the components in the following way:
\begin{align*}
(Tp)_i & = p_i-\epsilon \\
(Tp)_j & = p_j+\epsilon
  ~,
\end{align*}
where $0 < \epsilon < p_i-p_j$, while leaving all other components equal;
$(Tp)_k = p_k$ for $k\neq i,j$.
\end{Def}

Intuitively, these operations reduce concentration, since they act to equalize
the disparity between two components, in such a way as to not create greater
disparity in the opposite direction. This is the \emph{principle of transfers}.

Suppose now that we have two vectors $\mathbf{p}=(p_i)$ and $\mathbf{q}=(q_i)$
and that there exists a sequence of transfer operations
$\mathbf{T}_1,\dots,\mathbf{T}_m$ such that $\mathbf{T}_m \circ \cdots \circ
\mathbf{T}_1 \mathbf{p} = \mathbf{q}$. We will say that $\mathbf{p}$
\emph{majorizes} $\mathbf{q}$; denoted $\mathbf{p}\succsim \mathbf{q}$. The
relation $\succsim$ defines a \emph{preorder} on the set of distributions, as
it is reflexive and transitive but not necessarily antisymmetric.

There are, in fact, a number of equivalent criteria for majorization. We list
three relevant to our development in the following composite theorem.

\begin{The}[Majorization Criteria]
Given two vectors $\mathbf{p}=(p_i)$ and $\mathbf{q}=(q_i)$ with the same total sum, 
let their orderings be given by the permuted vectors
$\mathbf{p}^{\downarrow} = (p^{\downarrow}_i)$ and $\mathbf{q}^{\downarrow} = (q^{\downarrow}_i)$
such that $p^{\downarrow}_1 > p^{\downarrow}_2 > \dots > p^{\downarrow}_n$ and the same for $\mathbf{q}^{\downarrow}$.
Then the following statements are equivalent:
\begin{enumerate}
\setlength{\topsep}{0pt}
\setlength{\itemsep}{0pt}
\setlength{\parsep}{0pt}
\item \emph{Hardy-Littlewood-P\'olya:} For every $1\leq k\leq n$,
	\begin{align*}
	\sum_{i=1}^k p_i^{\downarrow} \geq \sum_{i=1}^k q_i^{\downarrow}
	~;
	\end{align*}
\item \emph{Principle of transfers:} $\mathbf{p}$ can be transformed
	to $\mathbf{q}$ via a sequence of transfer operations;
\item \emph{Schur-Horn:} There exists a unitary matrix $\mathbf{U}=(U_{ij})$
	such that $\mathbf{q} = \mathbf{Dp}$, where
	$\mathbf{D}=\left(\left|U_{ij}\right|^2\right)$,
	a \emph{uni-stochastic} matrix.
\end{enumerate}
\end{The}

\begin{figure}
\centering
\includegraphics[width=\columnwidth]{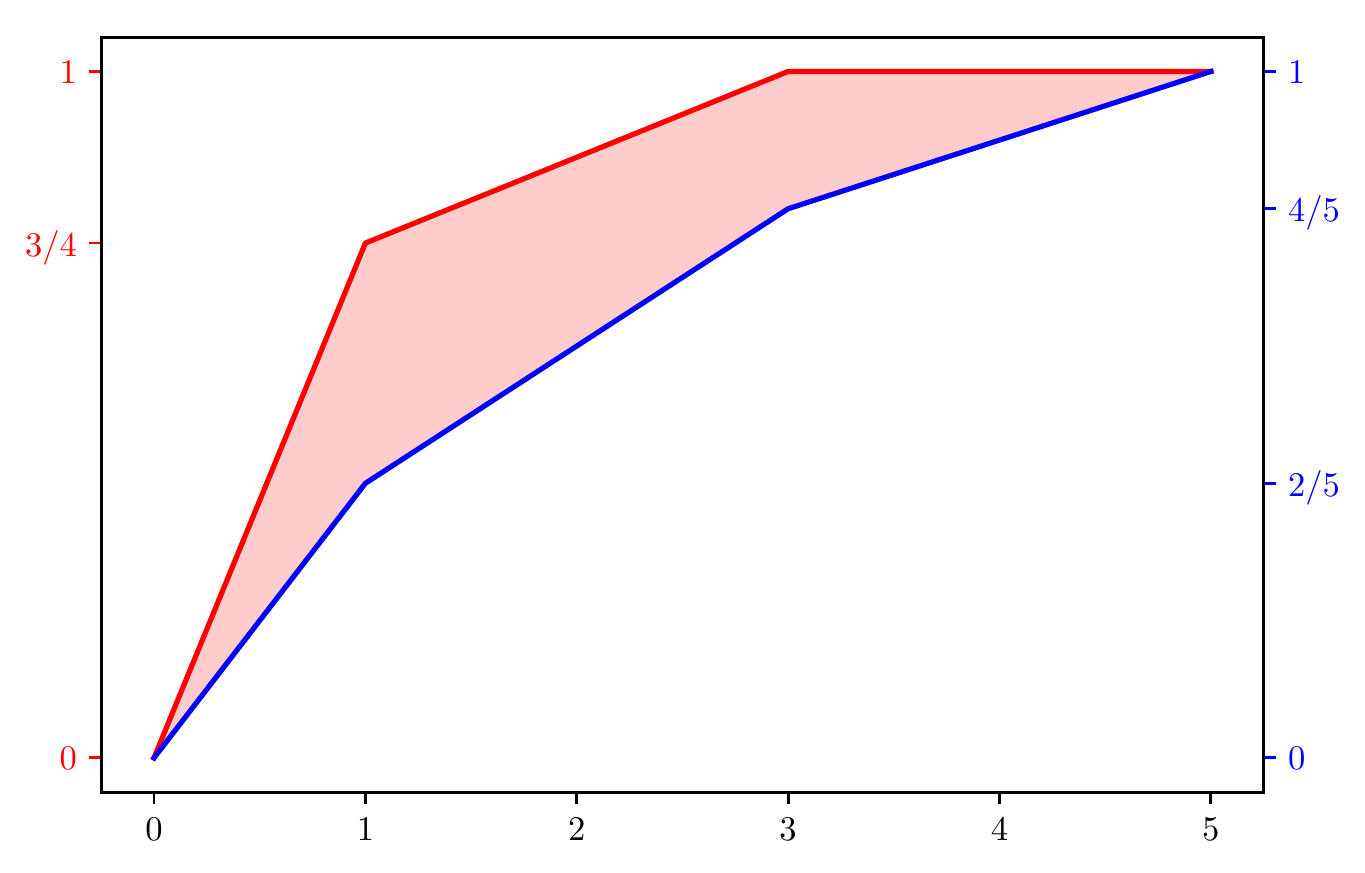}
\caption{{\color{red}$\mathbf{p}$} and {\color{blue}$\mathbf{q}$} are
	comparable and the first majorizes the second: $\mathbf{p} \succsim
	\mathbf{q}$. (Here we chose {$\mathbf{p} =
	(\nicefrac{3}{4},\nicefrac{1}{8},\nicefrac{1}{8},0,0)$} and {$\mathbf{q} =
	(\nicefrac{2}{5},\nicefrac{1}{5},\nicefrac{1}{5},\nicefrac{1}{10},\nicefrac{1}{10})$}.
	Tick marks indicate kinks in the Lorenz curve.)
	}
\label{fig:Lorenz_Compare}
\end{figure}

The Hardly-Littlewood-P\'olya criterion provides a visual representation of
majorization in the form of the \emph{Lorenz curve}. For a distribution
$\mathbf{p}=(p_i)$, the Lorenz curve is simply the function
$\beta_\mathbf{p}(k) = \sum_{i=1}^k p_i^{\downarrow}$. See Fig.
\ref{fig:Lorenz_Compare}. We can see that $\mathbf{p} \succsim \mathbf{q}$ so
long as the area under $\beta_\mathbf{q}$ is completely contained in the area
under $\beta_\mathbf{p}$.

The Lorenz curve can be understood via a social analogy, by examining rhetoric
of the form ``The top $x$\% of the population owns $y$\% of the wealth''. Let
$y$ be a function of $x$ in this statement, and we have the Lorenz curve of a
wealth distribution. (Majorization, in fact, has its origins in the study of
income inequality.)

\begin{figure}
\centering
\includegraphics[width=\columnwidth]{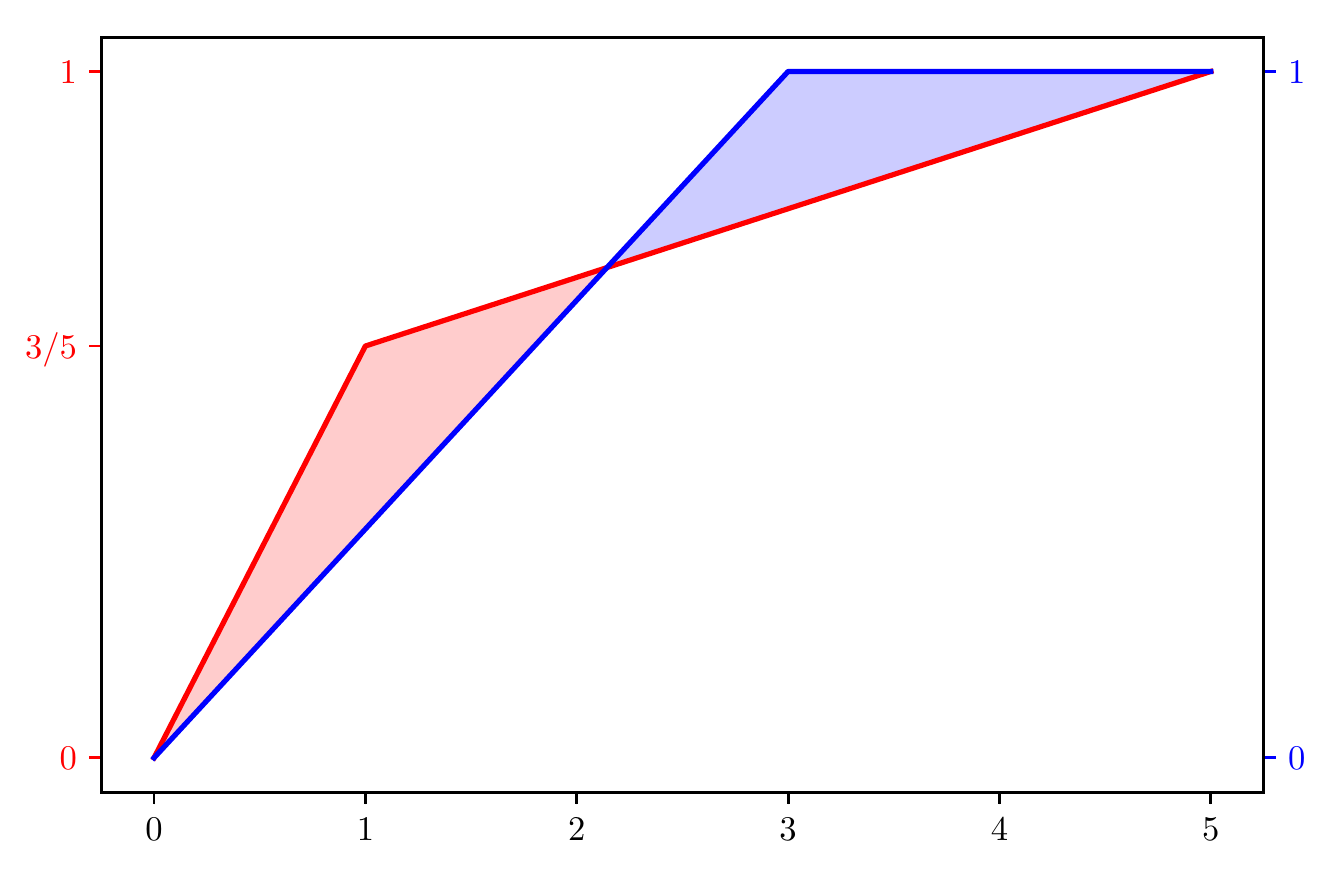}
\caption{{\color{red}$\mathbf{p}$} and {\color{blue}$\mathbf{q}$} are incomparable. (Here we chose {$\mathbf{p} = (\nicefrac{3}{5},\nicefrac{1}{10},\nicefrac{1}{10},\nicefrac{1}{10},\nicefrac{1}{10})$} and {$\mathbf{q} = (\nicefrac{1}{3},\nicefrac{1}{3},\nicefrac{1}{3},0,0)$}.)
	}
\label{fig:Lorenz_Incompare}
\end{figure}

If neither $\mathbf{p}$ nor $\mathbf{q}$ majorizes the other, they are \emph{incomparable}. (See Fig. \ref{fig:Lorenz_Incompare}.)

As noted, majorization is a preorder, since there may exist distinct
$\mathbf{p}$ and $\mathbf{q}$ such that $\mathbf{p}\succsim \mathbf{q}$ and
$\mathbf{q} \succsim \mathbf{p}$. This defines an equivalence relation $\sim$
between distributions. Every preorder can be converted into a partial order by
considering equivalence classes $[\mathbf{p}]_\sim$. 

If majorization, in fact, captures important physical properties of the
distributions, we should expect that these properties may be quantified. The
class of monotones that quantify the preorder of majorization are called
\emph{Schur-convex} and \emph{Schur-concave} functions.

\begin{Def}[Schur-convex (-concave) functions]
A function $f:\mathbb{R}^n\rightarrow\mathbb{R}$ is called Schur-convex
(-concave) if $\mathbf{p}\succsim \mathbf{q}$ implies $f(\mathbf{p}) \geq
f(\mathbf{q})$ ($f(\mathbf{p}) \leq f(\mathbf{q})$).
\end{Def}

An important class of Schur-concave functions consists of the R\'enyi entropies:
\begin{align*}
H_\alpha(\mathbf{p}) = \frac{1}{1-\alpha}\log_2\left(\sum_{i=1}^n p_i^\alpha\right)
  ~.
\end{align*}
In particular, the three limits
\begin{align*}
H(\mathbf{p}) & = \lim_{\alpha\rightarrow 1}H_\alpha(\mathbf{p})  = -\sum_{i=1}^n p_i\log_2 p_i
  ~,\\
H_0(\mathbf{p}) & = \lim_{\alpha\rightarrow 0}H_\alpha(\mathbf{p}) = \log_2 \left|\{1\leq i \leq n: p_i > 0\}\right|
  ~, ~\text{and} \\
H_\infty(\mathbf{p}) & = \lim_{\alpha\rightarrow \infty}H_\alpha(\mathbf{p}) 
  = -\log_2 \max_{1\leq i \leq n}p_i
  ~,
\end{align*}
---\emph{Shannon} entropy, \emph{topological} entropy, and
\emph{min}-entropy, respectively---describe important practical features of a
distribution. In order, they describe the asymptotic rate at which the outcomes
can be accurately conveyed, the single-shot resource requirements for the same
task, and the probability of error in guessing the outcome if no information is
conveyed at all (or, alternatively, the single-shot rate at which randomness
can be extracted from the distribution) \cite{Renn04a,Toma12a}. As such, they
play a significant role in communication and memory storage.

\begin{figure}
\centering
\includegraphics[width=\columnwidth]{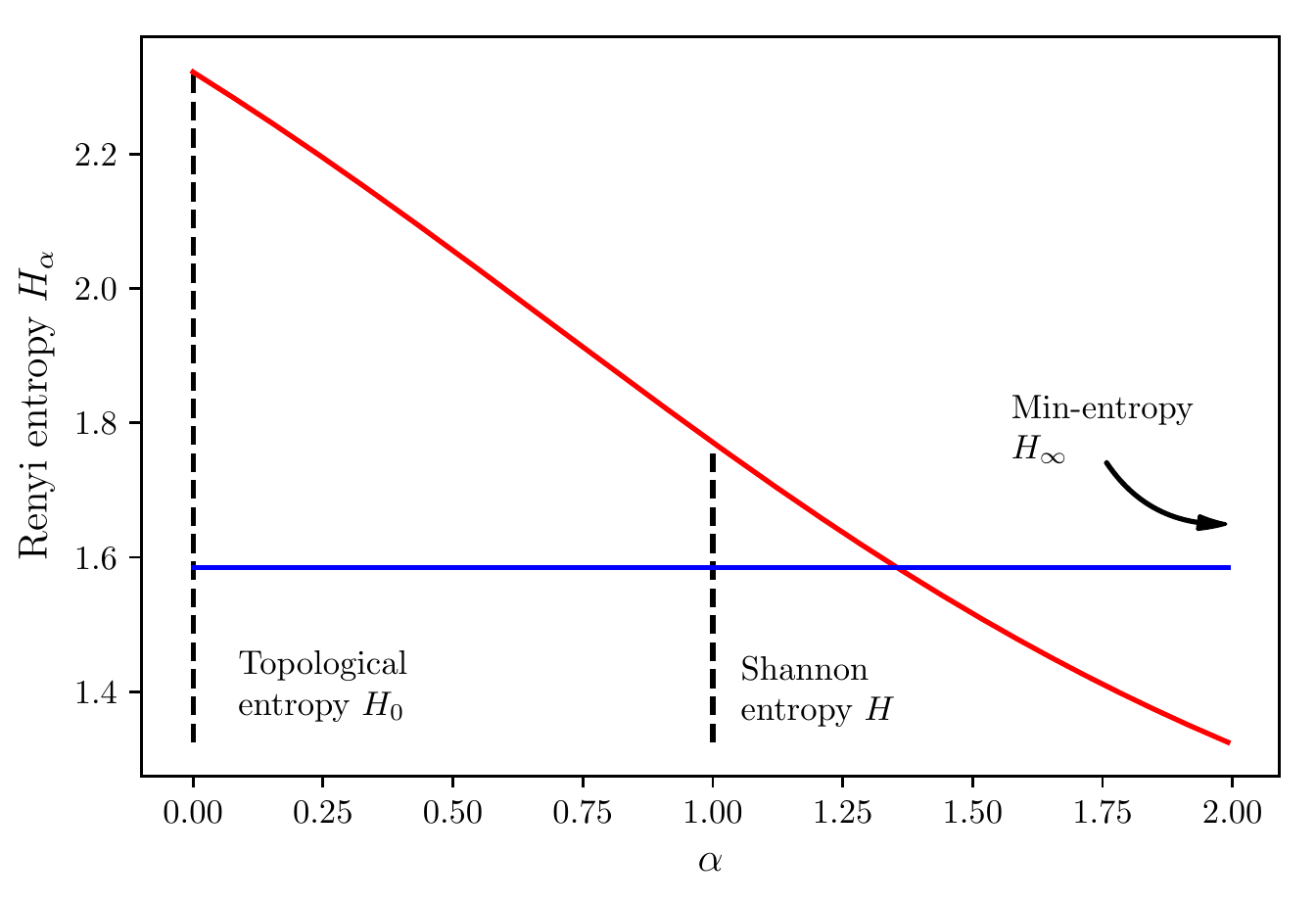}
\caption{R\'enyi entropies of the two incomparable distributions
	{\color{red}$\mathbf{p}$} and {\color{blue}$\mathbf{q}$} from Fig.
	\ref{fig:Lorenz_Incompare}.
	}
\label{fig:RenyiIncompare}
\end{figure}

The example of two incomparable distributions $\mathbf{p}$ and $\mathbf{q}$ can
be analyzed in terms of the R\'enyi entropies if we plot
$H_\alpha\left(\mathbf{p}\right)$ and $H_\alpha\left(\mathbf{q}\right)$ as a
function of $\alpha$, as in Fig. \ref{fig:RenyiIncompare}.

The central question we explore in the following is applying majorization to
determine when it is possible to simultaneously optimize all entropy monotones
or, alternatively, to determine if each monotone has a unique solution. This
leads to defining \emph{strong maxima} and \emph{strong minima}.

\begin{Def}[Strong maximum (minimum)]
Let $S$ be a set of probability distributions. If a distribution $\mathbf{p}\in
S$ satisfies $\mathbf{p}\precsim\mathbf{q}$ ($\mathbf{p}\succsim\mathbf{q}$),
for all $\mathbf{q}\in S$, then $\mathbf{p}$ is a \emph{strong maximum}
(\emph{minimum}) of the set $S$.
\end{Def}

The extrema names derive from the fact that the strong maximum maximizes the
R\'enyi entropies and the strong minimum minimizes them. One can extend the
definitions to the case where $\mathbf{p} \not\in S$, but is the
least-upper-bound such that any other $\mathbf{p}'$ satisfying
$\mathbf{p}'\precsim\mathbf{q}$ must obey $\mathbf{p}'\precsim\mathbf{p}$.
This case would be called a \emph{strong supremum} (or in the other direction a
\emph{strong infimum}). However, these constructions may not be unique as
$\succsim$ is a preorder and not a partial order. However, if we sort by
equivalence class, then the strongly maximal (minimal) class is unique if it
exists.

In lattice-theoretic terms, the strong maximum is essentially the lattice-theoretic notion of a \emph{meet}
and the strong minimum is a \emph{join} \cite{Grat10a}.

One example of strong minimization is found in quantum mechanics. Let $\rho$ be
a state and $X$ be a maximal diagonalizing measurement. For a given measurement
$Y$, let $\left.\rho\right|_Y$ be the corresponding probability distribution
that comes from measuring $\rho$ with $Y$. Then $\left.\rho\right|_X \succsim
\left.\rho\right|_Y$ for all maximal projective measurements $Y$. (This follows
from the unitary matrices that transform from the basis of $X$
to that of $Y$, and the Schur-Horn lemma.)

Another, recent example is found in Ref. \cite{Horo18a}, where the set
$B_{\epsilon}\left(\mathbf{p}\right)$ of all distributions $\epsilon$-close to
$\mathbf{p}$ under the total variation distance $\delta$ is considered:
\begin{align*}
B_{\epsilon}\left(\mathbf{p}\right)
  = \left\{\mathbf{q}:\delta(\mathbf{p},\mathbf{q})\leq \epsilon\right\}
  ~.
\end{align*}
This set has a strong minimum, called the \emph{steepest distribution}
$\overline{\mathbf{p}}{}_\epsilon$, and a strong maximum, called the
\emph{flattest distribution} $\underline{\mathbf{p}}{}_\epsilon$.

When a strong minimum or maximum does not exist, we refer to the individual extrema
of the various monotones as \emph{weak} extrema.

We close with a technical note on how to compare distributions over different
numbers of events. There are generally two standards for such comparisons that
depend on application. In the resource theory of informational nonequilibrium
\cite{Gour15a}, one compares distributions over different numbers of events
by ``squashing'' their Lorenz curves so that the $x$-axis ranges from $0$ to
$1$. Under this comparison, the distribution $\mathbf{p}_3 = (1,0,0)$ has more
informational nonequilibrium than $\mathbf{p}_2=(1,0)$. In the following,
however, we adopt the standard of simply extending the smaller distribution by
adding events of zero probability. In this, $\mathbf{p}_3$ and $\mathbf{p}_2$
are considered equivalent. This choice is driven by our interest in the R\'enyi
entropic costs and not in the overall nonequilibrium. (The latter is more
naturally measured by R\'enyi \emph{negentropies}
$\bar{H}_\alpha\left(\mathbf{p}\right) = \log n -
{H}_\alpha\left(\mathbf{p}\right)$, where $n$ is the number of events.)

\section{Strong minimality of the \EM}
\label{sec:EMUniversality}

The general task we set ourselves is simulating classical processes.

\begin{Def}[Bi-infinite process]
A bi-infinite process over an alphabet $\mathcal{A}$ is a probability measure
$\mathbb{P}(\overleftrightarrow{x})$ over the set of all bi-infinite strings
$\overleftrightarrow{x}= \overleftarrow{x}_t \overrightarrow{x}_t \in
\mathcal{A}^{\infty}$, where the past $\overleftarrow{x}_t = \dots x_{-1+t}
x_t$ and the future $\overrightarrow{x}_t = x_t x_{t+1} \dots$ are constructed
by concatenating elements of $\mathcal{A}$.
\end{Def}

Though defined over bi-infinite strings, the measure gives probabilities for
seeing finite-length words $w=x_1\dots x_\ell$, defined as $\mathbb{P}(w) =
\mathbb{P}\left(\{ \overleftrightarrow{x} = \overleftarrow{x}_t w
\overrightarrow{x}_{t+\ell} : t \in \mathbb{N}\}\right)$. This can be taken as an alternate
definition of the process measure.

Here, we focus on finite predictive models.

\begin{Def}[Finite predictive model]
A \emph{finite predictive model} is a triplet $\mathbf{M}
=(\mathcal{R},\mathcal{A},\{\mathbf{T}^{(x)}:x\in\mathcal{A}\})$ of hidden
states $\mathcal{R}$, an output alphabet $\mathcal{A}$, and nonnegative
transition matrices $\mathbf{T}^{(x)} = \left(T^{(x)}_{\rho\rho'} \right)$ with
$x\in\mathcal{A}$ and $\rho,\rho'\in\mathcal{R}$, satisfying the properties:
\begin{enumerate}
\setlength{\topsep}{0pt}
\setlength{\itemsep}{0pt}
\setlength{\parsep}{0pt}
\item \emph{Irreducibility}:
	$\mathbf{T} = \sum_{x\in\mathcal{A}} \mathbf{T}^{(x)}$ is stochastic and
	irreducible.
\item \emph{Unifilarity}:
	$T^{(x)}_{\rho\rho'} = \mathbb{P}\left(x\middle|\rho\right)
	\delta_{\rho',f(\rho,x)}$ for some conditional probability
	$\mathbb{P}\left(x\middle|\rho\right)$ and deterministic function $f$.
\end{enumerate}
\end{Def}

A finite predictive model is a type of hidden Markov model \cite{Uppe97a},
whose dynamic is to transition between states at each timestep while emitting a
symbol with probabilities determined by the transition matrices
$\mathbf{T}^{(x)}$. Unifilarity ensures that, given the model state $\sigma \in
\mathcal{R}$ and symbol $x \in \mathcal{A}$, the next state $\sigma^\prime \in
\mathcal{R}$ is unique.

Given a finite predictive model $\mathbf{M}$, the state transition matrix
$\mathbf{T}$ has a single left-eigenstate $\boldsymbol{\pi}$ of eigenvalue $1$,
by the Perron-Frobenius theorem, satisfying $\boldsymbol{\pi}^\top \mathbf{T} =
\boldsymbol{\pi}^\top$. We call this state distribution the \emph{stationary
state}. Using it, we define the process $\mathbb{P}_\mathbf{M}$ generated by
$\mathbf{M}$ as $\mathbb{P}_\mathbf{M}(w) = \boldsymbol{\pi}^\top
\mathbf{T}^{(x_1)}\cdots\mathbf{T}^{(x_\ell)} \boldsymbol{1}$, where
$w=x_1\dots x_\ell$ and $\boldsymbol{1}$ is the vector with all 1's for its
components. $\mathbb{P}_\mathbf{M}$ describes a stationary process. If we let
$\boldsymbol{\delta}_\rho$ represent the state-distribution that assigns the
state $\rho\in\mathcal{R}$ probability $1$, then
$\mathbb{P}_{\mathbf{M},\rho}(w) = \boldsymbol{\delta}_{\rho}^\top
\mathbf{T}^{(x_1)}\cdots\mathbf{T}^{(x_\ell)} \boldsymbol{1}$ is the
probability of seeing word $w$ after starting in state $\rho$.

Given a model with stationary distribution $\boldsymbol{\pi}$, we define the
model's R\'enyi memory as $H_\alpha\left(\mathbf{M}\right) =
H_\alpha\left(\boldsymbol{\pi}\right)$. This includes the topological memory
$H_0\left(\mathbf{M}\right)$, the statistical memory
$H\left(\mathbf{M}\right)=H_1\left(\mathbf{M}\right)$, and the min-memory
$H_\infty\left(\mathbf{M}\right)$. Given a process $\mathbf{P}$, we define the
R\'enyi \emph{complexity} as the minimal memory $\Cmu^{(\alpha)} =
\min_{\mathbf{M}}H_\alpha\left(\mathbf{M}\right)$ over all models that generate
$\mathbf{P}$ \cite{Crut88a}. These include the topological complexity
$C^{(0)}_\mu$, the statistical complexity $\Cmu=\Cmu^{(1)}$, and the
min-complexity $C^{(\infty)}_\mu$.

Among the class of finite predictive models, a particularly distinguished
member is the \emph{\eM} \cite{Crut88a}:
\begin{Def}[Generator \eM]
A generator \eM is a finite predictive model $\mathbf{M}
=(\mathcal{S},\mathcal{A},\{\mathbf{T}^{(x)}:x\in\mathcal{A}\})$ such that for
each pair of distinct states $\rho, \rho' \in \mathcal{S}$, there exists a
word $w$ such that $\mathbb{P}_{\mathbf{M},\rho}(w)
\neq\mathbb{P}_{\mathbf{M},\rho'}(w)$.
\end{Def}
In other words, a generator \eM must be irreducible, unifilar, and its states
must be {\em probabilistically distinct}, so that no pair of distinct states
predict the same future.

An important result of computational mechanics is that the generator \eM is
unique with respect to the process it generates \cite{Trav12a}. This is a
combined consequence of the equivalence of the generator definition with
another, called the \emph{history} \eM, which is provably unique
\cite{Shal98a}. That is, given an \eM $\mathbf{M}$, there is no other \eM that
generates $\mathbb{P}_{\mathbf{M}}$. A further important result is that the \eM
minimizes both the statistical complexity $\Cmu$ and the topological complexity
$\Cmu^{(0)}$.

\begin{figure}
\centering
(a) \includegraphics[width=0.5\columnwidth]{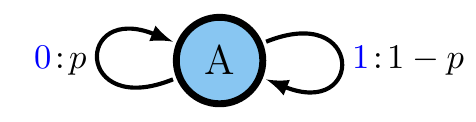} \\
(b) \includegraphics[width=0.8\columnwidth]{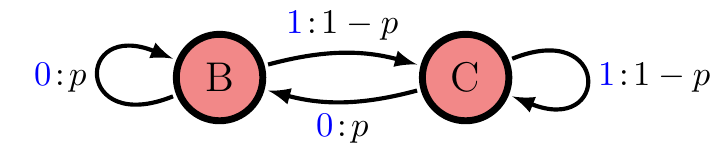} \\
(c) \includegraphics[width=0.75\columnwidth]{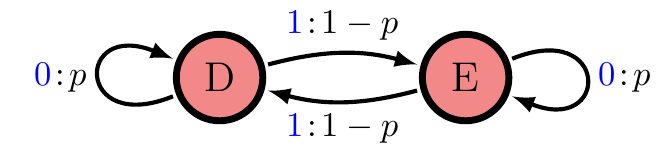}
\caption{(a) \EM for a coin flipped with bias $p$. (b) Alternate representation
	with $p$ to be in state $B$ and $1-p$ to be in state $C$. (c) Alternate
	representation with biases $p$ to stay in current state and $1-p$ to switch
	states.
	}
\label{fig:machines}
\end{figure}

To fix intuitions, consider now several examples of models and their processes.
First, consider the Biased Coin Process, a memoryless process in which, at each
time step, a coin is flipped with probability $p$ of generating a $1$ and
probability $1-p$ of generating a $0$. Figure \ref{fig:machines} displays three
models for it. Model (a) is the process' \eM, and models (b) and (c) are each
$2$-state alternative finite predictive models. Notice that in both models (b)
and (c), the two states generate equivalent futures.

\begin{figure}
\centering
(a) \includegraphics[width=0.45\columnwidth]{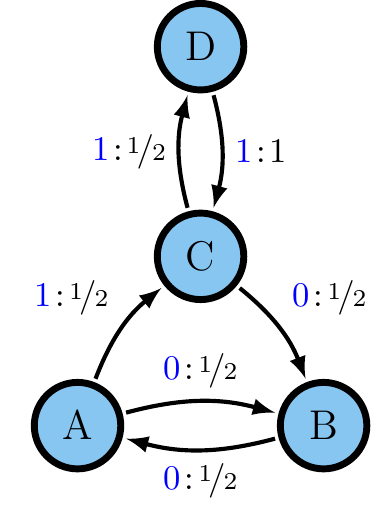} \\
(b) \includegraphics[width=0.75\columnwidth]{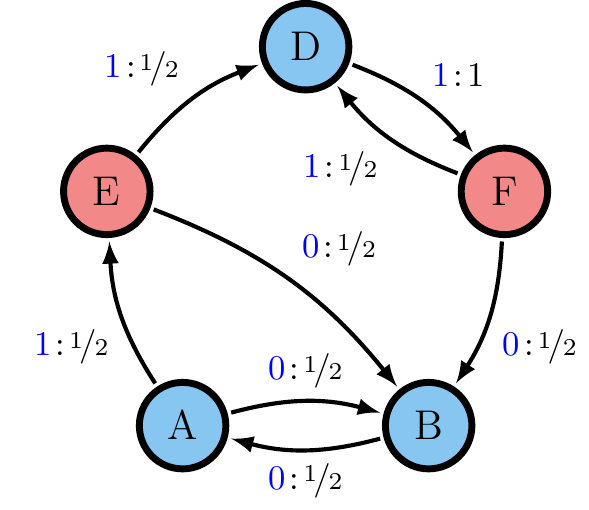}
\caption{(a) \EM for Even-Odd Process. (b) Refinement of the Even-Odd Process
	\eM, where the \eM's state $C$ has been split into states $E$ and $F$.
	}
\label{fig:EvenOddMachines}
\end{figure}

Continuing, Fig. \ref{fig:EvenOddMachines} displays two alternative models of
the Even-Odd Process. This process produces sequences formed by concatenating
strings of an odd number of $1$s to strings of an even number of $0$s. We see
in (a) the process' \eM. In (b), we see an alternative finite predictive model,
and notice that its states $E$ and $F$ predict the same futures, and so are not
probabilistically distinct. We notice that they both play the role of state $C$
in the \eM, in terms of the futures they predict.

\begin{figure}
\centering
(a) \includegraphics[width=\columnwidth]{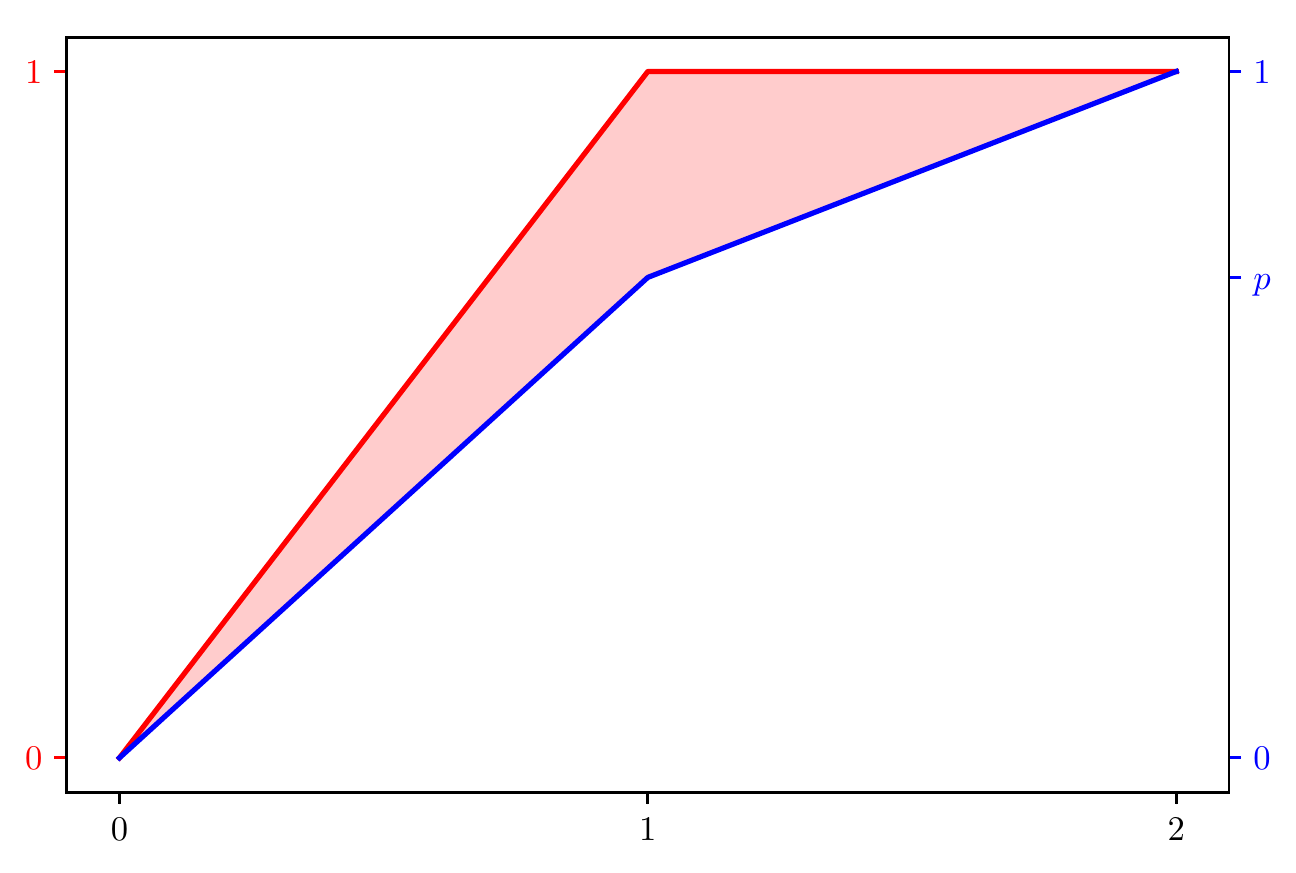} \\
(b) \includegraphics[width=\columnwidth]{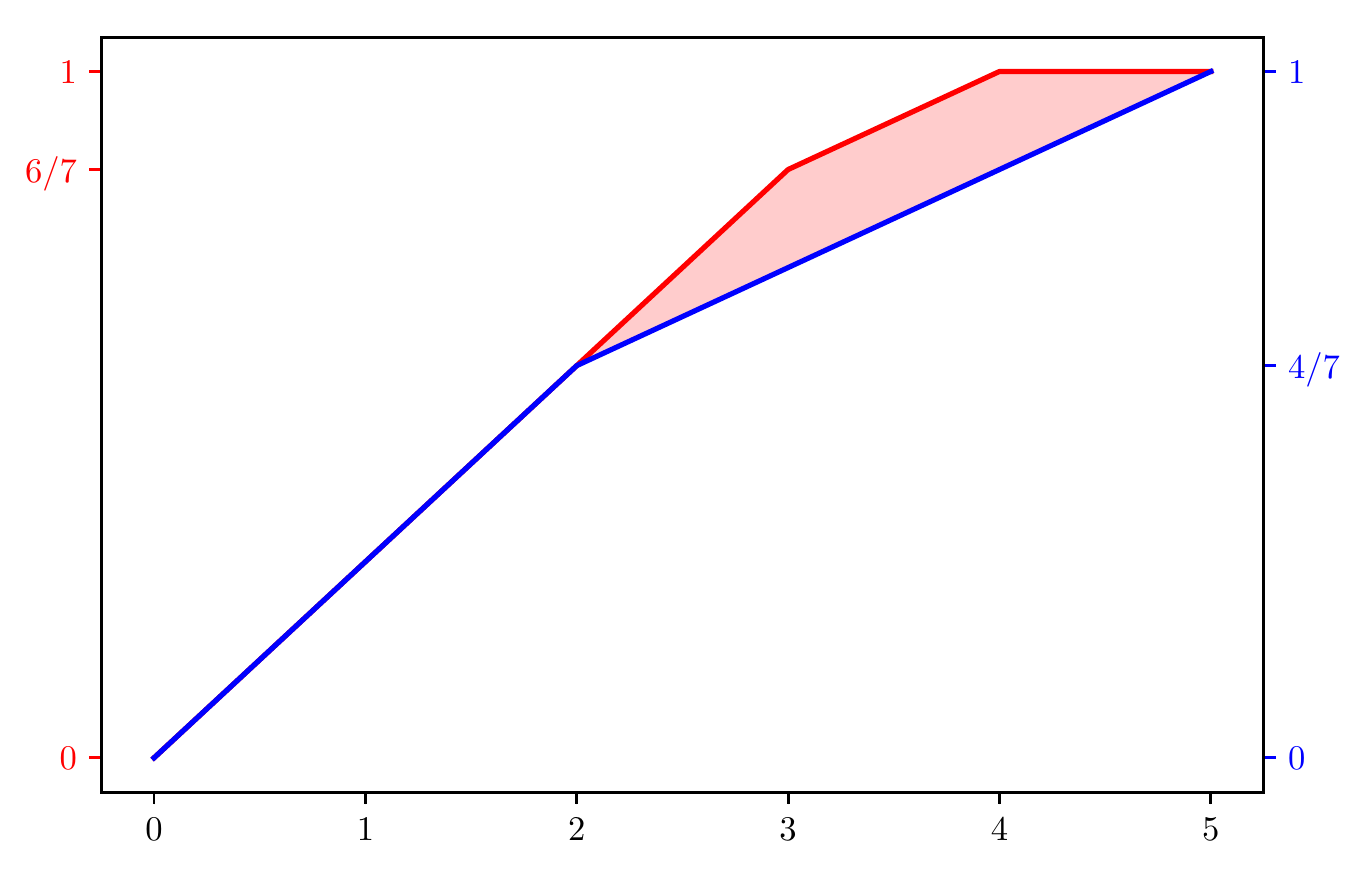}
\caption{(a) Lorenz curves for Fig.
	\ref{fig:machines}{\color{red} \textbf{(a)}}'s \eM and Fig.
	\ref{fig:machines}{\color{blue} \textbf{(b)}}'s alternative predictor of
	the Biased Coin Process. (b) Same comparison for the Even-Odd Process
	\eM Fig. \ref{fig:EvenOddMachines}{\color{red} \textbf{(a)}} and
	alternative predictor Fig. \ref{fig:EvenOddMachines}{\color{blue}
	\textbf{(b)}}.
	}
\label{fig:LorenzMachineCompare}
\end{figure}

We can compare these examples using Lorenz curves of the state distributions,
as shown in Fig. \ref{fig:LorenzMachineCompare}. Here, recall, we adopted the
convention of comparing two distributions over a different number of states by
extending the smaller system to include zero-probability states. We notice that
the \eM state distribution always majorizes the state distribution of the
alternative machines.

The key to formalizing this observation is the following lemma.

\begin{Lem}[State Merging]
\label{lem:StateMerg}
Let $\mathbf{M}=(\mathcal{R},\mathcal{A},\{\mathbf{T}^{(x)}:x\in\mathcal{A}\})$
be a finite predictive model that is not an \eM. Then the machine created by
merging its probabilistically equivalent states is the \eM of the process
$\mathbb{P}_\mathbf{M}$ generated by $\mathbf{M}$.
\end{Lem}

\begin{ProLem}
Let $\sim$ be the equivalence relation $\rho\sim\rho'$ if
$\mathbb{P}_{\mathbf{M},\rho}(w) =\mathbb{P}_{\mathbf{M},\rho'}(w)$ for all
$w$. Let $\mathcal{S}$ consist of the set of equivalence classes
$[\rho]_{\sim}$ generated by this relation. For a given class $\sigma\in
\mathcal{S}$, consider the transition probabilities associated with each
$\rho\in\sigma$. For each $x\in\mathcal{A}$ such that
$\mathbb{P}\left(x\middle|\rho\right) > 0$, there is a outcome state $\rho_x =
f(x,\rho)$. Comparing with another state in the same class $\rho'\in\sigma$, we
have the set of outcome states $\rho_x' = f(x,\rho')$. For the future
predictions of both states $\rho$ and $\rho'$ to be equivalent, they must also
be equivalent after seeing the symbol $x$. That is,
$\mathbb{P}_{\mathbf{M},\rho}(w) =\mathbb{P}_{\mathbf{M},\rho'}(w)$ for all $w$
also implies $\mathbb{P}_{\mathbf{M},\rho}(xw)
=\mathbb{P}_{\mathbf{M},\rho'}(xw)$ for all $w$. But
$\mathbb{P}_{\mathbf{M},\rho}(xw) = \mathbb{P}_{\mathbf{M},\rho_x}(w)$, and so
we have $\rho_x\sim \rho_x'$ for all $x\in\mathcal{A}$.

The upshot of these considerations is that we can define a consistent and
unifilar transition dynamic
$\{\widetilde{\mathbf{T}}{}^{(x)}:x\in\mathcal{A}\}$ on $\mathcal{S}$ given by
the matrices $\widetilde{T}{}^{(x)}_{\sigma\sigma'} =
\widetilde{T}{}^{(x)}_{\rho\rho'}$ for any $\rho\in\sigma$ and
$\rho'\in\sigma'$.  It inherits unifilarity from the original model
$\mathbf{M}$ as well as irreducibility. It has probabilistically distinct
states because we have already merged all of the probabilistically equivalent
states. Therefore, the resulting machine $\mathbf{M}_{\mathcal{S}} =
(\mathcal{S},\mathcal{A},\{\widetilde{\mathbf{T}}{}^{(x)}:x\in\mathcal{A}\})$
is the \eM of the process $\mathbb{P}_\mathbf{M}$ generated by $\mathbf{M}$.
\end{ProLem}

The state-merging procedure here is an adaptation of the Hopcroft algorithm for
minimization of deterministic (nonprobabilistic) finite automata, which is
itself an implementation of the Nerode equivalence relation, \cite{Hopc71a}.
It has been applied previously to analyze synchronization in \eMs
\cite{Trav11a}.

Using Lemma \ref{lem:StateMerg}, we can prove the main result of this section:

\begin{The}[Strong Minimality of \EM]
Let $\mathbf{M}_{\mathcal{S}} =
(\mathcal{S},\mathcal{A},\{\widetilde{\mathbf{T}}{}^{(x)}:x\in\mathcal{A}\})$
be the \eM of process $\mathbb{P}$ and
$\mathbf{M}_{\mathcal{R}}=(\mathcal{R},\mathcal{A},\{\mathbf{T}^{(x)}:x\in\mathcal{A}\})$
be any other finite generating machine. Let the stationary distributions be
$\boldsymbol{\pi}_\mathcal{S}=\left(\pi_{\mathcal{S},\sigma}\right)$ and
$\boldsymbol{\pi}_\mathcal{R}=\left(\pi_{\mathcal{R},\rho}\right)$,
respectively. Then $\boldsymbol{\pi}_\mathcal{S}\succsim
\boldsymbol{\pi}_\mathcal{R}$.
\end{The}

\begin{ProThe}
By Lemma \ref{lem:StateMerg}, the states of the \eM $\mathbf{M}_{\mathcal{S}}$
are formed by merging equivalence classes $\sigma=[\rho]$ on the finite
predictive model $\mathbf{M}_{\mathcal{R}}$. Since the machines are otherwise
equivalent, the stationary probability $\pi_{\mathcal{S},\sigma}$ is simply the
sum of the stationary probabilities for each $\rho\subseteq\sigma$, given by
$\pi_{\mathcal{R},\rho}$. That is:
\begin{align*}
\pi_{\mathcal{S},\sigma}
  = \sum_{\rho\in \mathcal{R}_\sigma} \pi_{\mathcal{R},\rho}
  ~.
\end{align*}
One can then construct $\boldsymbol{\pi}_\mathcal{R}$ from
$\boldsymbol{\pi}_\mathcal{S}$ by a series of transfer operations in which
probability is shifted out of the state $\sigma$ into new states $\rho$. Since
the two states are related by a series of transfer operations,
$\boldsymbol{\pi}_\mathcal{S}\succsim \boldsymbol{\pi}_\mathcal{R}$.
\end{ProThe}

It immediately follows from this that not only does the \eM minimize the
statistical complexity $\Cmu$ and the topological complexity $C^{(0)}_\mu$, but
it also minimizes every other R\'enyi complexity $C^{(\alpha)}_\mu$ as well.

The uniqueness of the \eM is extremely important in formulating this result.
This property of \eMs\ follows from the understanding of predictive models as
partitions of the past and of the \eMs\ as corresponding to the coarsest
graining of these predictive partitions \cite{Shal98a}. Other paradigms for
modeling will not necessarily have this underlying structure and so may not
have strongly minimal solutions. In the following, we see this is, in fact, the
case for pure-state quantum machines.

\section{Strong quantum advantage}
\label{sec:Ambiguity}

A pure-state quantum model can be generalized from the classical case by
replacing the classical states $\sigma$ with quantum-mechanical pure states
$\left|\eta_\sigma\right>$ and the symbol-labeled transition matrices
$\mathbf{T}^{(x)}$ with symbol-labeled Kraus operators $K^{(x)}$.

\begin{Def}[Pure-state quantum model]
A pure-state quantum model is a quintuplet $\mathbf{M}
=(\mathcal{H},\mathcal{A},\mathcal{S},\Sigma=\{\left|\eta_\sigma\right>:\sigma\in \mathcal{S}\},\{K^{(x)}:x\in\mathcal{A}\})$
of a Hilbert space $\mathcal{H}$, an output alphabet $\mathcal{A}$, pure states
$\left|\eta_\sigma\right>$ corresponding to some set of state labels
$\mathcal{S}$, and nonnegative Kraus operators $K^{(x)}$ with
$x\in\mathcal{A}$ satisfying the properties:
\begin{enumerate}
\setlength{\topsep}{0pt}
\setlength{\itemsep}{0pt}
\setlength{\parsep}{0pt}
\item \emph{Completeness}: The Kraus operators satisfy
	$\sum_x K^{(x)\dagger}K^{(x)}=I$.
\item \emph{Unifilarity}: $K^{(x)}\left|\eta_\sigma\right> \propto \left|\eta_{f(\sigma,x)}\right>$ for some deterministic function $f(\sigma,x)$.
\end{enumerate}
\end{Def}

This is a particular kind of hidden quantum Markov model \cite{Monr12a} in
which we assume the dynamics can be described by the evolution of pure states.
This is practically analogous to the assumption of unifilarity in the classical
predictive setting.

It is not necessarily the case that the states $\{\left|\eta_\sigma\right>\}$
form an orthonormal basis; rather, nonorthonormality is the intended advantage
\cite{Gu12a,Maho15a}. Overlap between the states allows for a smaller von
Neumann entropy for the stationary state of the process. We formalize this
shortly.

It is assumed that the Kraus operators have a unique stationary state
$\rho_\pi$. One way to compute it is to note that taking $\mathbb{P}(x|\sigma)
= \left<\eta_\sigma\right|K^{(x)\dagger}K^{(x)}\left|\eta_\sigma\right>$ and
the function $\sigma \mapsto f(\sigma,x)$ determines a finite predictive model
as defined above. The model's stationary state $\boldsymbol{\pi}=(\pi_\sigma)$
is related to the stationary state of the quantum model via:
\begin{align*}
\rho_\pi = \sum_{\sigma} \pi_\sigma
\left|\eta_\sigma\right>\left<\eta_\sigma\right|
  ~.
\end{align*}
The process generated by a pure-state quantum model has the word distribution,
for words $w=x_1 \dots x_\ell$:
\begin{align*}
\mathbb{P}(w) = \mathrm{Tr}\left[K^{(x_\ell)}\cdots K^{(x_1)}\rho_\pi K^{(x_1)\dagger}\cdots K^{(x_L)\dagger}\right]
  ~.
\end{align*}

The eigenvalues $\{\lambda_i\}$ of the stationary state $\rho_\pi$ form
a distribution $\boldsymbol{\lambda} = \left(\lambda_i\right)$. 
The R\'enyi entropies of these distributions form the {\em von Neumann-R\'enyi} entropies
of the states:
\begin{align*}
S_\alpha\left(\rho_\pi\right) = H_\alpha \left(\boldsymbol{\lambda}\right)
  ~.
\end{align*}
We noted previously that for a given state these are strongly minimal over the
entropies of all projective, maximal measurements on the state. Given a model
$\mathbf{M}$ with stationary state $\rho_\pi$, we may simply write
$S_\alpha\left(\mathbf{M}\right) = S_\alpha\left(\rho_\pi\right)$ as the
R\'enyi memory of the model. Important limits, as before, are the topological
memory $S_0\left(\mathbf{M}\right)$, the statistical memory
$S\left(\mathbf{M}\right) = S_1\left(\mathbf{M}\right)$, and the min-memory
$S_\infty\left(\mathbf{M}\right)$, which represent physical limitations on
memory storage for the generator.

To properly compare pure-state quantum models and classical predictive models,
we define the {\em classical equivalent model} of a pure-state quantum model.

\begin{Def}[Classical equivalent model]
Let $\mathbf{M}
=(\mathcal{H},\mathcal{A},\mathcal{S},\Sigma=\{\left|\eta_\sigma\right>:\sigma\in
\mathcal{S}\},\{K^{(x)}:x\in\mathcal{A}\})$ be a pure-state quantum model, with
probabilities and deterministic function $\mathbb{P}(x|\sigma) =
\left<\eta_\sigma\right|K^{(x)\dagger}K^{(x)}\left|\eta_\sigma\right>$ and
$\sigma \mapsto f(\sigma,x)$, respectively. Its classical equivalent
$\mathbf{M}_\mathrm{Cl.}$ is the classical finite predictive model with state
set $\mathcal{S}$, alphabet $\mathcal{A}$ and symbol-based transition matrices
$\mathbf{T}^{(x)}$ generated by the state-to-symbol probabilities
$\mathbb{P}(x|\sigma)$ and deterministic function $f(\sigma,x)$.
\end{Def}

We now prove that a finite classical predictive model strongly maximizes
all pure-state quantum models of which it is the classical equivalent.

\begin{The}[Strong quantum advantage]
\label{StrongQAdvantage}
Let $\mathbf{M}
=(\mathcal{H},\mathcal{A},\mathcal{S},\Sigma=\{\left|\eta_\sigma\right>:\sigma\in
\mathcal{S}\},\{K^{(x)}:x\in\mathcal{A}\})$ be a pure-state quantum model with
stationary state $\rho_\pi$, and let $\mathbf{M}_\mathrm{Cl.}$ be the classical
equivalent model with stationary state $\boldsymbol{\pi}=(\pi_\sigma)$ (with
$\sigma=1,\dots,n$). Let $d=\dim \mathcal{H}$ and $n=
\left|\mathcal{S}\right|$. (We have $n\geq d$: if not, then we can take a
smaller Hilbert space that spans the states.) Let
$\boldsymbol{\lambda}=(\lambda_i)$ be an $n$-dimensional vector where the first
$d$ components are the eigenvalues of $\rho_\pi$ and the remaining elements are
$0$. Then $\boldsymbol{\lambda}\succsim \boldsymbol{\pi}$.
\end{The}

\begin{ProThe}
We know that:
\begin{align*}
\rho_\pi & = \sum_{\sigma\in\mathcal{S}} \pi_\sigma
     \left|\eta_\sigma\right>\left<\eta_\sigma\right| \\
  & = \sum_{\sigma\in\mathcal{S}}
     \left|\phi_\sigma\right>\left<\phi_\sigma\right|
  ~,
\end{align*}
where $\left|\phi_\sigma\right> = \sqrt{\pi_\sigma}\left|\eta_\sigma\right>$.
However, we can also write $\rho_\pi$ in the eigenbasis:
\begin{align*}
\rho_\pi & = \sum_{i=1}^d \lambda_i \left|i\right>\left<i\right| \\
  & = \sum_{i=1}^d  \left|\psi_i\right>\left<\psi_i\right|
  ~,
\end{align*}
where $\left|\psi_i\right> = \sqrt{\lambda_i}\left|i\right>$. Then the two sets
of vectors can be related via:
\begin{align*}
\left|\phi_\sigma\right> = \sum_{i=1}^d U_{\sigma i} \left|\psi_i\right>
  ~,
\end{align*}
where $U_{\sigma i}$ is a $n\times d$ matrix comprised of $d$ rows of
orthonormal $n$-dimensional vectors \cite{Hugh93}. Now, we have:
\begin{align*}
\pi_\sigma & = \left<\phi_\sigma|\phi_\sigma\right> \\
  & = \sum_{i=1}^d |U_{\sigma i}|^2 \lambda_i
  ~.
\end{align*}
Note that $U_{\sigma i}$ is not square, but since we have taken $\lambda_i = 0$
for $i> d$, we can simply extend $U_{\sigma i}$ into a square unitary matrix by
filling out the bottom $n-d$ rows with more orthonormal vectors. This
leaves the equation unchanged. We can then write:
\begin{align*}
\pi_\sigma = \sum_{i=1}^n |U_{\sigma i}|^2 \lambda_i
  ~.
\end{align*}
Then by Theorem 1, $\boldsymbol{\lambda}\succsim \boldsymbol{\pi}$.
$\square$
\end{ProThe}

\begin{Cor}
$S_\alpha(\mathbf{M}) \leq H_\alpha\left(\mathbf{M}_{\mathrm{Cl.}}\right)$ for all $\alpha \geq 0$.
\label{cor:QRenyi}
\end{Cor}

\begin{ProCor}
$S_\alpha(\rho_\pi) \leq H_\alpha(\boldsymbol{\pi})$ for all $\alpha \geq 0$
follows from the definitions of the von Neumann-R\'enyi entropies and the
Schur-concavity of $H_\alpha$.
$\square$
\end{ProCor}

Many alternative pure-state quantum models may describe the same process.  The
``first mark'', so to speak, for quantum models is the $q$-machine
\cite{Maho15a,Riec15b}, which directly embeds the dynamics of the \eM\ into a
quantum system while already leveraging the memory advantage due to state
overlap.

\begin{Def}[$q$-Machine]
Given an \eM\
$\{\mathcal{S},\mathcal{A},\{\mathbf{T}^{(x)}:x\in\mathcal{A}\}\}$, where
${T}^{(x)}_{\sigma\sigma'}=\mathbb{P}(x|\sigma) \delta_{\sigma',f(\sigma,x)}$
for some deterministic function $f(\sigma,x)$, construct the corresponding
$q$-machine in the following way:
\begin{enumerate}
\setlength{\topsep}{0pt}
\setlength{\itemsep}{0pt}
\setlength{\parsep}{0pt}
\item The states $\left|\eta_\sigma\right>$ are built to satisfy the
recursive relation:
\begin{align*}
\left<\eta_\sigma|\eta_{\sigma'}\right> = \sum_{x\in\mathcal{A}}\sqrt{\mathbb{P}(x|\sigma)\mathbb{P}(x|\sigma')}\left<\eta_{f(\sigma,x)}|\eta_{f(\sigma',x)}\right>
  ~.
\end{align*}
\item $\mathcal{H}$ is the space spanned by the states
	$\left|\eta_\sigma\right>$.
\item The Kraus operators $K^{(x)}$ are determined by the relations:
\begin{align*}
K^{(x)}\left|\eta_\sigma\right> = \sqrt{\mathbb{P}(x|\sigma)}\left|\eta_{f(\sigma,x)}\right>
  ~.
\end{align*}
\end{enumerate}
\end{Def}

One can check that this satisfies the completeness relations and has the
correct probability dynamics for the process generated by the \eM.

That the $q$-machine offers statistical memory advantage with respect to the
\eM was previously shown in \cite{Maho16a} and with respect to topological
memory in \cite{Thom18a}. Theorem \ref{StrongQAdvantage} and Corollary
\ref{cor:QRenyi} imply these as well as advantage with respect to other R\'enyi
measures of memory.

\section{Weak quantum minimality}
\label{sec:WeakMin}

An open problem is to determine the minimal quantum pure-state representation
of a given classical process. This problem is solved in some specific instances
such as the Ising model \cite{Suen17a} and the Perturbed Coin Process
\cite{Thom18a}. In these cases it is known to be the $q$-machine. We denote the
smallest value of the R\'enyi entropy of the stationary state as
$C_{q}^{(\alpha)} = \min_{\mathbf{M}} S_\alpha\left(\mathbf{M}\right)$, called
the \emph{quantum R\'enyi complexities}, including the limits, the quantum
topological complexity $C_{q}^{(0)}$, the quantum min-complexity
$C_{q}^{(\infty)}$, and the quantum statistical complexity $C_{q}=C_{q}^{(1)}$.
If a strongly minimal quantum pure-state model exists, these complexities are
all attained by the same pure-state model. One of our primary results in this
section is that for some processes, this does not occur.

\begin{figure}
\centering
\includegraphics[width=0.95\columnwidth]{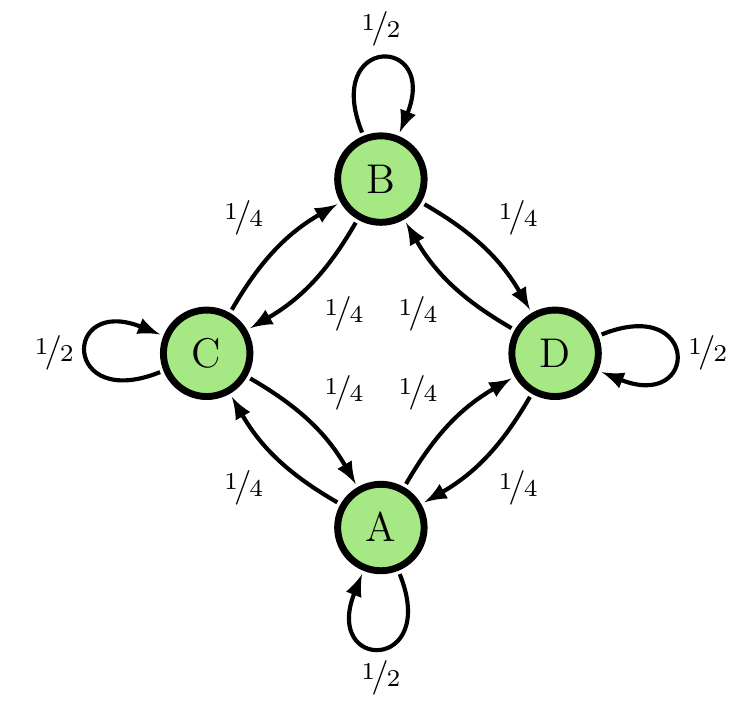}
\caption{The 4-state MBW Process as a Markov chain (which is the \eM).}
\label{fig:4Monras}
\end{figure}

We start by examining two examples. The first, the MBW Process introduced in
Ref. \cite{Monr12a}, demonstrates a machine whose $q$-machine is not minimal in
the von Neumann complexity. Consider the process generated by the \emph{4-state
MBW} machine shown in Fig. \ref{fig:4Monras}.

This process' HMM is simply a Markov chain, and its representation in Fig. \ref{fig:4Monras} is its \eM.
Denote this classical representation by $\mathbf{M}_4$.
If we take $\{ \left|A\right>, \left|B\right>,
\left|C\right>, \left|D\right> \}$ as an orthonormal basis of a Hilbert space,
we can construct the $q$-machine with the states:
\begin{align*}
\left|\eta_A\right> & = \frac{1}{\sqrt{2}}\left|A\right> +
\frac{1}{2}\left(\left|C\right>+\left|D\right>\right) ~, \\
\left|\eta_B\right> & = \frac{1}{\sqrt{2}}\left|B\right> +
\frac{1}{2}\left(\left|C\right>+\left|D\right>\right) ~, \\
\left|\eta_C\right> & = \frac{1}{\sqrt{2}}\left|C\right> +
\frac{1}{2}\left(\left|A\right>+\left|B\right>\right) ~, ~\text{and}\\
\left|\eta_D\right> & = \frac{1}{\sqrt{2}}\left|D\right> + \frac{1}{2}\left(\left|A\right>+\left|B\right>\right)
  ~.
\end{align*}
Since it is a Markov chain, we can write the Kraus operators as $K_x =
\left|\eta_x\right>\left<\epsilon_x\right|$, where
$\left<\epsilon_x|\eta_{x'}\right> \propto \sqrt{\mathbb{P}(x|x')}$. This is a
special case of the construction used in Ref. \cite{Agha18a}. For $q$-machines
of Markov chains, then, the dual basis is just $\left<\epsilon_x\right| =
\left<x\right|$. We denote the $q$-machine model of the 4-state MBW Process as
$\mathbf{Q}_4$.

\begin{figure}
\centering
\includegraphics[width=\columnwidth]{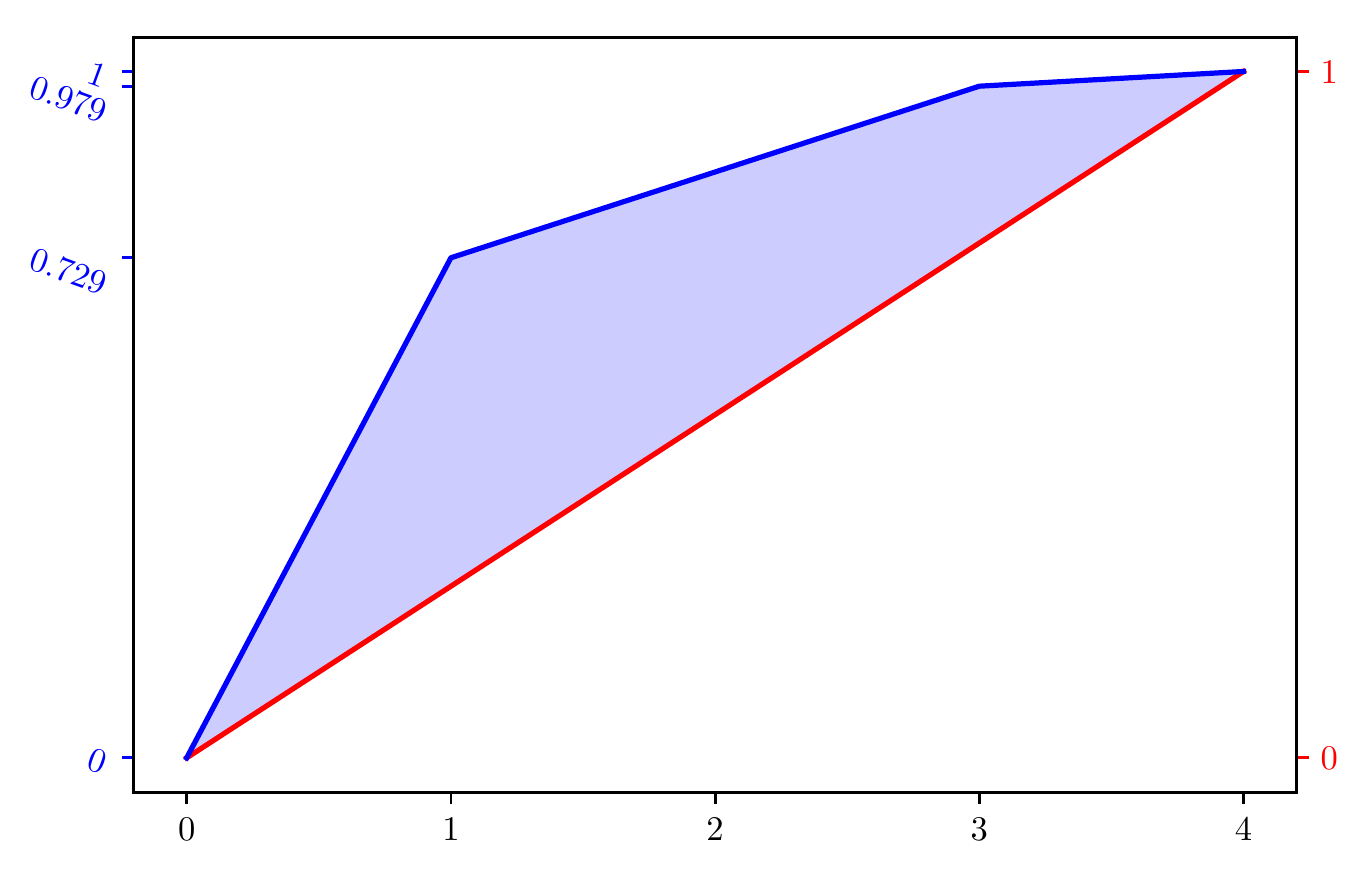}
\caption{Lorenz curves for the 4-state MBW \eM\ {\color{red}$\mathbf{M}_4$}
	and the associated $q$-machine {\color{blue}$\mathbf{Q}_4$}.
	}
\label{fig:Lorenz4MonrasQM}
\end{figure}

Let's examine the majorization between $\mathbf{Q}_4$ and the Markov model via
the Lorenz curves of $\boldsymbol{\lambda}$, the eigenvalues of $\rho_\pi$, and
the stationary state of the Markov chain. See Fig. \ref{fig:Lorenz4MonrasQM}.

\begin{figure}
\centering
\includegraphics[width=\columnwidth]{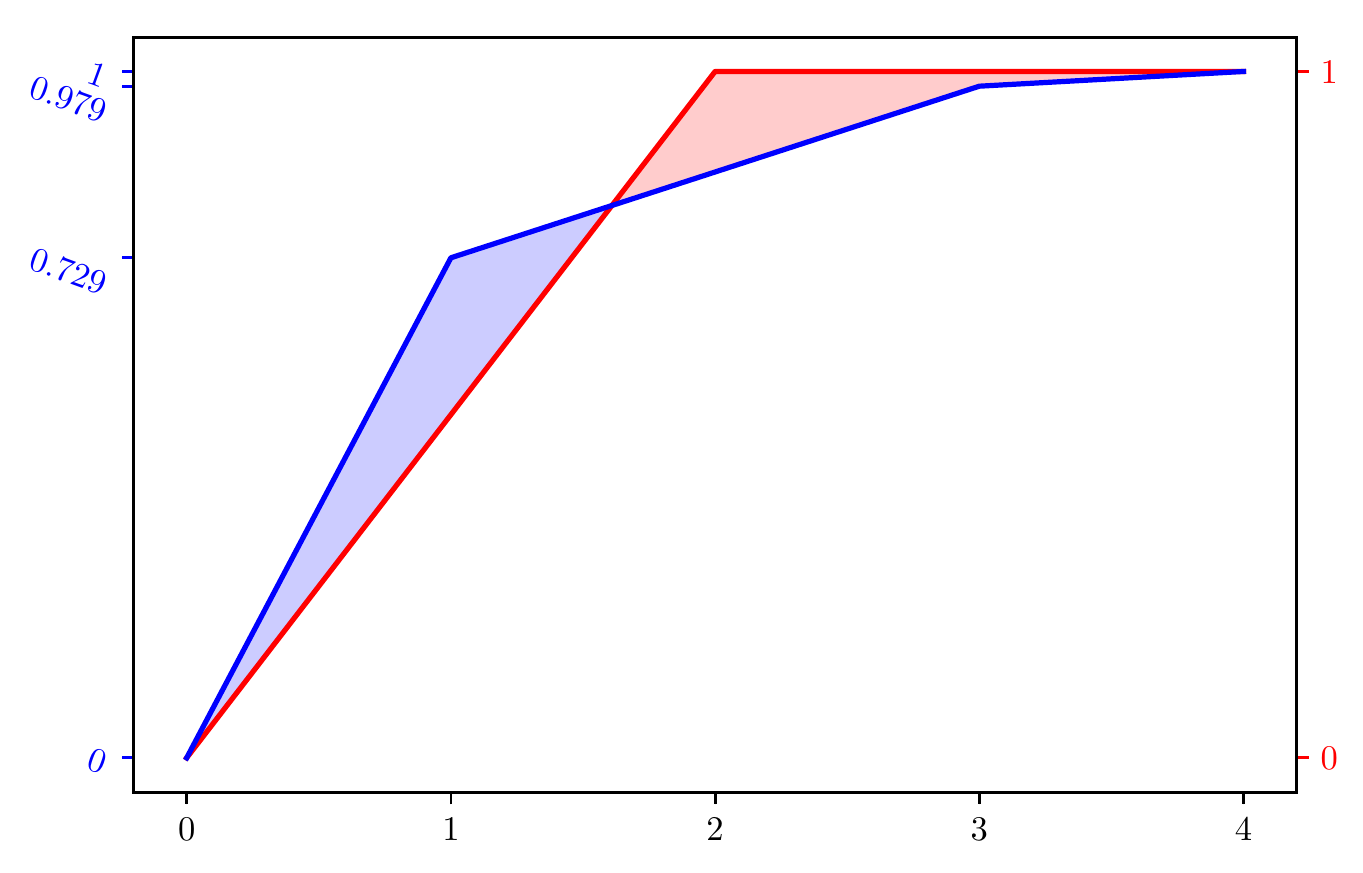}
\caption{Lorenz curves for the 4-state MBW $q$-machine
	{\color{blue}$\mathbf{Q}_4$} and a dimensionally smaller model
	{\color{red}$\mathbf{D}_4$}.
	}
\label{fig:Lorenz4MonrasSmallerModel}
\end{figure}

It turns out that there is a smaller quantum model embedded in two dimensions,
with states:
\begin{align*}
\left|\eta_A'\right> & = \left|0\right> ~, \\
\left|\eta_B'\right> & = \left|1\right> ~, \\
\left|\eta_C'\right> & =
\frac{1}{\sqrt{2}}\left(\left|0\right>+\left|1\right>\right) ~, ~\text{and}\\
\left|\eta_D'\right> & =
\frac{1}{\sqrt{2}}\left(\left|0\right>-\left|1\right>\right)
  ~.
\end{align*}
In this case, $\left<\epsilon_x'\right|=\frac{1}{\sqrt{2}}\left<\eta_x'\right|$
derives the $q$-machine. This gives the proper transition probabilities for the
4-state MBW model. This dimensionally smaller model we denote $\mathbf{D}_4$.
Figure \ref{fig:Lorenz4MonrasSmallerModel} compares the Lorenz
curve of its stationary eigenvalues $\boldsymbol{\lambda}'$ to those of $\mathbf{Q}_4$.
One sees that it does not majorize the $q$-machine, but it does have a lower
statistical memory: $S(\mathbf{D}_4) = 1.0$ and $S(\mathbf{Q}_4) \approx 1.2$ bit.
(On the other hand, the $q$-machine has a smaller min-memory, with
$S_\infty(\mathbf{D}_4) = 1.0$ and $S_\infty(\mathbf{Q}_4) \approx 0.46$.)

\begin{figure}
\centering
\includegraphics[width=0.65\columnwidth]{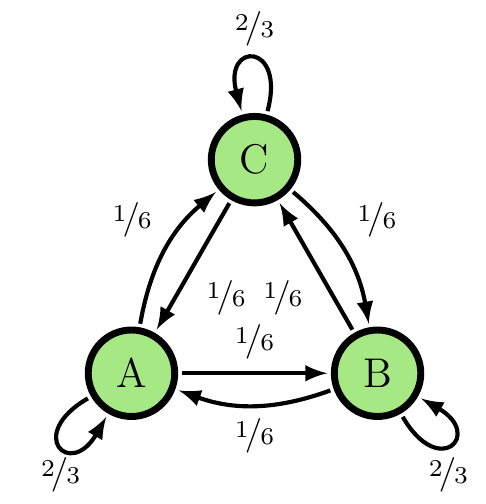}
\caption{3-state MBW Process as a Markov chain (which is the process' \eM).
	}
\label{fig:3Monras}
\end{figure}

Now consider something in the opposite direction. Consider the 3-state MBW
model, denoted $\mathbf{M}_3$ and displayed in Fig. \ref{fig:3Monras}. This is
a generalization of the previous example to three states instead of four.  We
will compute the corresponding $q$-machine $\mathbf{Q}_3$ and show that there
also exists a dimensionally smaller representation $\mathbf{D}_3$. In this
case, however, $\mathbf{D}_3$ is not smaller in its statistical memory.

\begin{figure}
\centering
\includegraphics[width=\columnwidth]{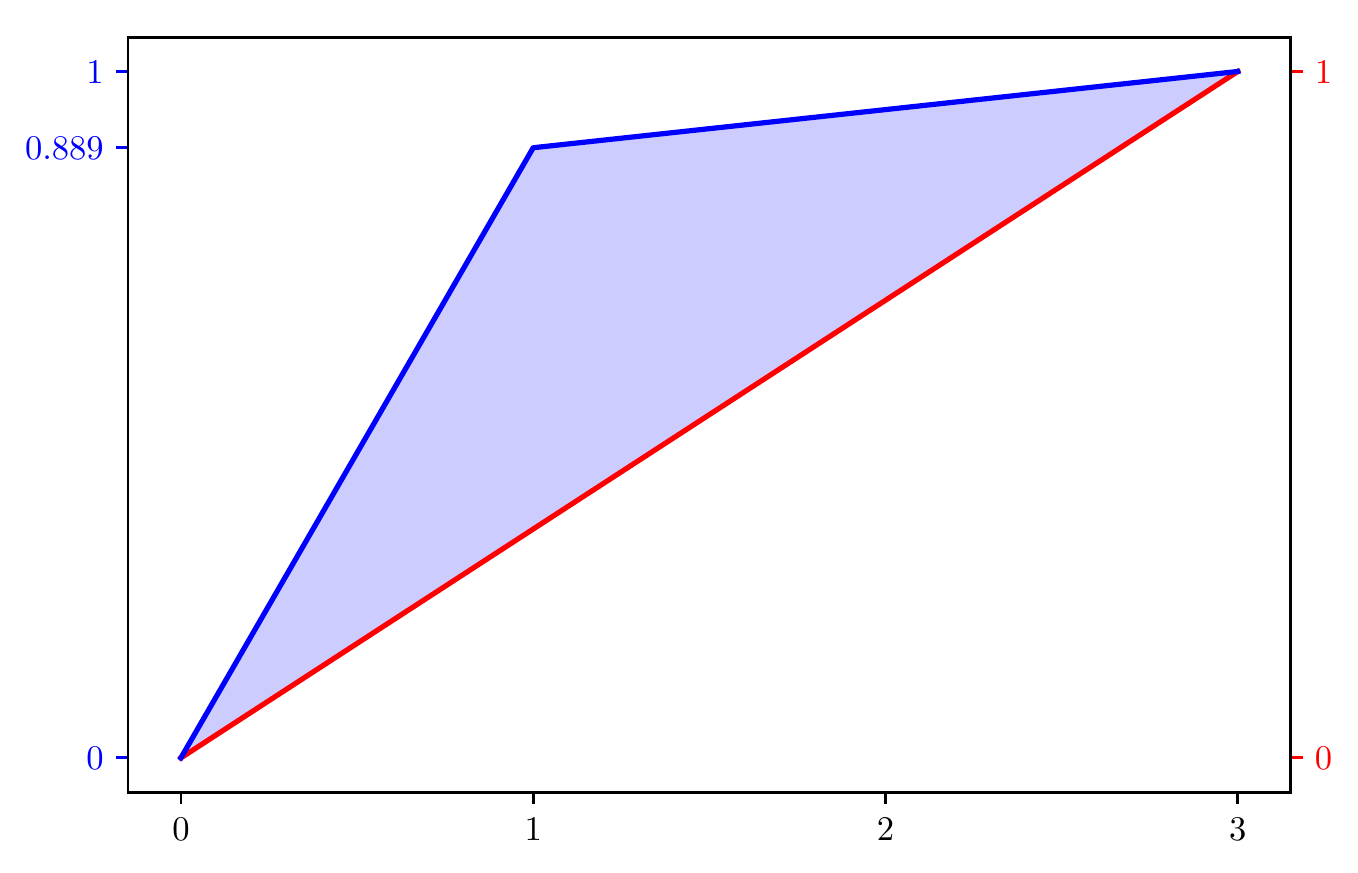}
\caption{Lorenz curves for the 3-state MBW \eM\ {\color{red}$\mathbf{M}_3$} and the associated
	$q$-machine {\color{blue}$\mathbf{Q}_3$}.
	}
\label{fig:3MonrasVQM}
\end{figure}

The $q$-machine $\mathbf{Q}_3$ of this Markov chain is given by the states:
\begin{align*}
\left|\eta_A\right> & = \sqrt{\frac{2}{3}}\left|A\right> +
\frac{1}{\sqrt{6}}\left(\left|B\right>+\left|C\right>\right) ~, \\
\left|\eta_B\right> & = \sqrt{\frac{2}{3}}\left|B\right> +
\frac{1}{\sqrt{6}}\left(\left|A\right>+\left|C\right>\right) ~, ~\text{and}\\
\left|\eta_C\right> & = \sqrt{\frac{2}{3}}\left|C\right> +
\frac{1}{\sqrt{6}}\left(\left|A\right>+\left|B\right>\right) ~,
\end{align*}
and Kraus operators defined similarly to before. We can examine the
majorization between the $q$-machine and the Markov model by plotting the
Lorenz curves of $\boldsymbol{\lambda}$, the eigenvalues of $\rho_\pi$, and the
stationary state of the Markov chain, shown in Fig. \ref{fig:3MonrasVQM}.

\begin{figure}
\centering
\includegraphics[width=\columnwidth]{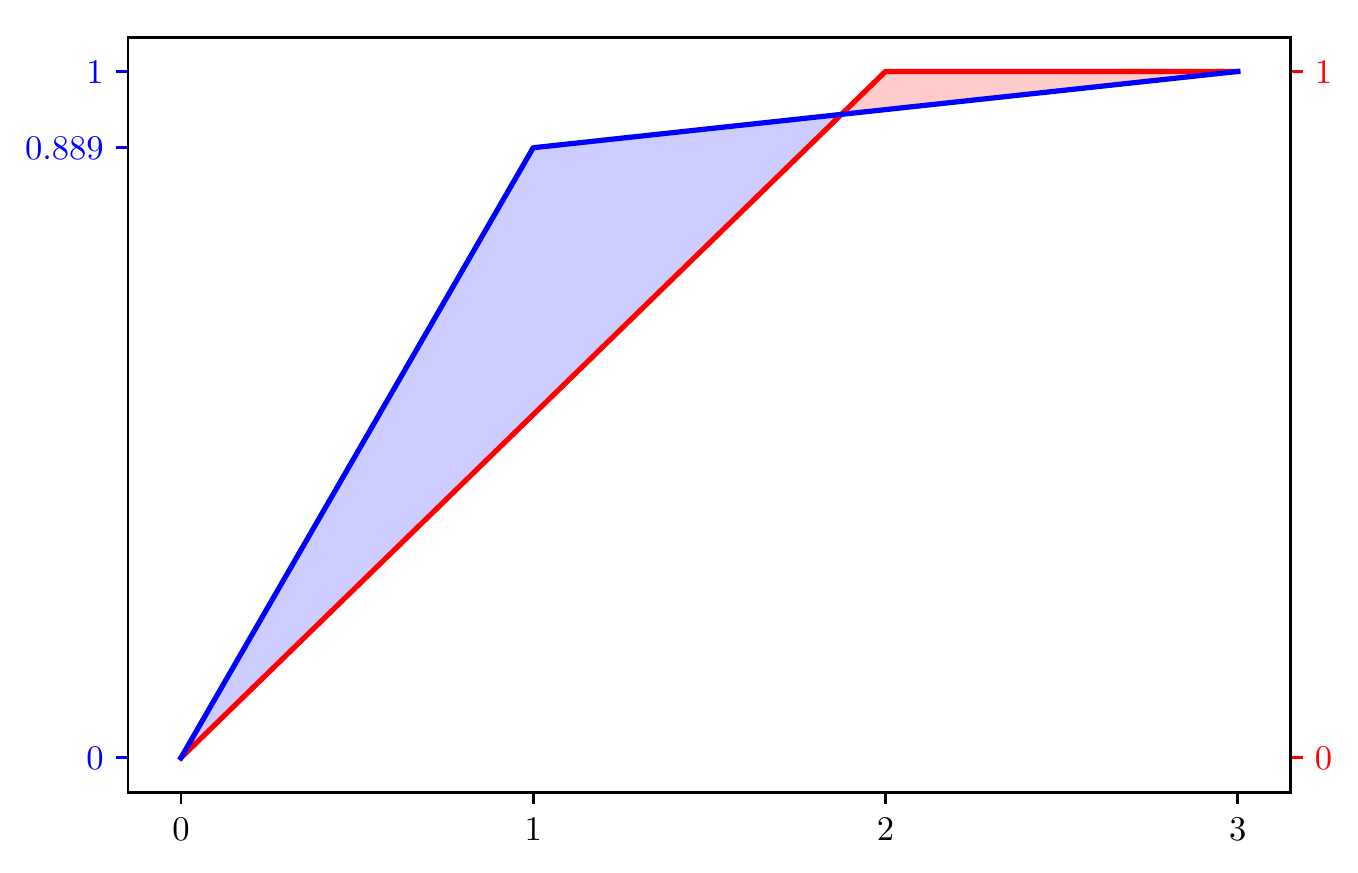}
\caption{Lorenz curves for the 3-state MBW $q$-machine,
	{\color{blue}$\mathbf{Q}_3$} and a
	dimensionally smaller model {\color{red}$\mathbf{D}_3$}.
	}
\label{fig:3MonrasVSmallerQM}
\end{figure}

The lower-dimensional model $\mathbf{D}_3$ is given by the states:
\begin{align*}
\left|\eta_A\right> & = \left|0\right> ~, \\
\left|\eta_B\right> & =
\frac{1}{2}\left|0\right>+\frac{\sqrt{3}}{2}\left|1\right> ~, ~\text{and}\\
\left|\eta_C\right> & = \frac{1}{2}\left|0\right>-\frac{\sqrt{3}}{2}\left|1\right>
  ~,
\end{align*}
with $\left<\epsilon_x'\right|=\sqrt{\frac{2}{3}}\left<\eta_x'\right|$. This
gives the proper transition probabilities for the 3-state MBW model. Figure
\ref{fig:3MonrasVSmallerQM} compares the Lorenz curve of its stationary
eigenvalues $\boldsymbol{\lambda}'$ to that of $\mathbf{Q}_3$. We see that it
does not majorize $\mathbf{Q}_3$. And, this time, this is directly manifested
by the fact that the smaller-dimension model has a larger entropy:
$S(\mathbf{D}_3) = 1.0$ and $S(\mathbf{Q}_3) \approx 0.61$ bit.

After seeing the \eM's strong minimality with respect to other classical models
and its strong maximality with respect to quantum models, it is certainly
tempting to conjecture that a strongly minimal quantum model exists. However,
the examples we just explored cast serious doubt. None of the examples covered
above are strong minima. One way to prove that no strong minimum exists for,
say, the 3-state MBW process requires showing that there does not exist
\emph{any other} quantum model in $2$ dimensions that generates the process.
This would imply that no other model can majorize $\mathbf{D}_3$. And, since
this model is not strongly minimal, no strongly minimal solution can exist.

Appendix \ref{app:QAmbiguity} proves exactly this---thus, demonstrating a
counterexample to the strong minimality of quantum models.

\begin{Count}[Weak Minimality of $\mathbf{D}_3$]
The quantum model $\mathbf{D}_3$ weakly minimizes topological complexity for
all quantum generators of the 3-state MBW Process; consequently, the 3-state
MBW Process has no strongly minimal model.
\end{Count}

\begin{figure}
\centering
\includegraphics[width=\columnwidth]{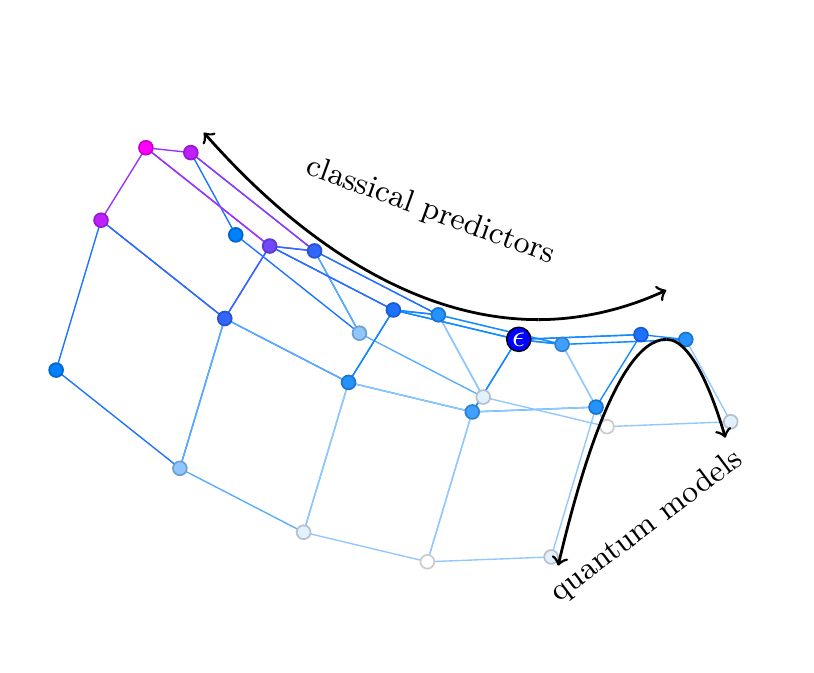}
\caption{Proposed majorization saddle structure of model-space: The \eM (labeled
	$\epsilon$) is located at a saddle-point with respect to majorization,
	where classical deviations (state-splitting) move up the lattice and
	quantum deviations (utilizing state overlap) move down the lattice.
	}
\label{fig:saddle}
\end{figure}

\section{Concluding remarks}
\label{sec:Conclude}

Majorizing states provides a means to compare a process' alternative
models in both the classical and quantum regimes. Majorization implies the
simultaneous minimization of a large host of functions. As a result we showed
that:
\begin{enumerate}
\setlength{\topsep}{0pt}
\setlength{\itemsep}{0pt}
\setlength{\parsep}{0pt}
\item The \eM\ majorizes all classical predictive models of the same process,
	and so simultaneously minimizes many different measures of memory cost.
\item The $q$-machine, and indeed any quantum realization of the \eM, always
	majorizes the \eM, and so simultaneously improves on all the measures of
	memory cost.
\item For at least one process, there does not exist any quantum pure-state
	model that majorizes all quantum pure-state models of that process. Thus,
	while an $\epsilon$-machine may be improved upon by different possible
	quantum models, there is not a unique one quantum model that is
	unambiguously the ``best'' choice.
\end{enumerate}
Imagining the \eM as an invariant ``saddle-point'' in the majorization
structure of model-space, Fig. \ref{fig:saddle} depicts the implied geometry.
That is, we see that despite its nonminimality among all models, the \eM still
occupies a topologically important position in model-space---one that is
invariant to one's choice of memory measure. However, no similar model plays
the topologically minimal role for quantum pure-state models.

The quantum statistical complexity $C_q$ has been offered up as an alternative
quantum measure of structural complexity---a rival of the statistical complexity
$\Cmu$ \cite{Tan14a}. One implication of our results here is that the nature of
this quantum minimum $C_q$ is fundamentally different than that of $\Cmu$. This
observation should help further explorations into techniques required to
compute $C_q$ and the physical circumstances in which it is most relevant. That
the physical meaning of $C_q$ involves generating an asymptotically large
number of realizations of a process may imply that it cannot be accurately
computed by only considering machines that generate a single realization. This
is in contrast to $\Cmu$ which, being strongly minimized, must be attainable in
the single-shot regime along with measures like $\Cmu^{(0)}$ and
$\Cmu^{(\infty)}$.

In this way, the quantum realm again appears ambiguous. Ambiguity in structural
complexity has been previously observed in the sense that there exist pairs of
processes, $A$ and $B$, such that $\Cmu(A)>\Cmu(B)$ but $C_q(A)<C_q(B)$
\cite{Agha17b}. The classical and quantum paradigms for modeling can disagree on
simplicity---there is no universal Ockham's Razor. How this result relates to
strong versus weak optimization deserves further investigation.

The methods and results here should also be extended to analyze classical
generative models which, in many ways, bear resemblances in their functionality
to the quantum models \cite{Loeh09a,Loeh09b,Rueb18a}. These drop the
requirement of unifilarity, similar to how the quantum models relax the notion
of orthogonality. Important questions to pursue in this vein are whether
generative models are strongly maximized by the \eM and whether they have their
own strong minimum or, like the quantum models, only weak minima in different
contexts.

To close, we only explored finite-state, discrete-time processes.
Processes with infinite memory \cite{Marz15a} and continuous generation
\cite{Marz17a,Elli18a} are also common in nature. Applying our results to
understand these requires further mathematical development.

\section*{Acknowledgments}
\label{sec:acknowledgments}

The authors thank Fabio Anza, John Mahoney, Cina Aghamohammadi, and Ryan James
for helpful discussions. As a faculty member, JPC thanks the Santa Fe Institute
and the Telluride Science Research Center for their hospitality during visits.
This material is based upon work supported by, or in part by, John Templeton
Foundation grant 52095, Foundational Questions Institute grant FQXi-RFP-1609,
the U.S. Army Research Laboratory and the U. S. Army Research Office under
contract W911NF-13-1-0390 and grant W911NF-18-1-0028, and via Intel Corporation
support of CSC as an Intel Parallel Computing Center.


\cleardoublepage

\appendix

\section{Appendix: Weak Minimality of $\mathbf{D}_3$}
\label{app:QAmbiguity}

Here, we prove that $\mathbf{D}_3$ is the unique 2D representation of the
3-state MBW process. We show this by considering the entire class of 2D models
and applying the completeness constraint.

We note that a pure-state quantum model of the $3$-state MBW process must have
three states $\left|\eta_A\right>$, $\left|\eta_B\right>$, and
$\left|\eta_C\right>$, along with three dual states $\left<\epsilon_A\right|$,
$\left<\epsilon_B\right|$, and $\left<\epsilon_C\right|$ such that: 
\begin{align*}
\left<\epsilon_A|\eta_A\right> & = e^{i\phi_{AA}}\sqrt{\frac{2}{3}} ~,\\
\left<\epsilon_A|\eta_B\right> & = e^{i\phi_{AB}}\frac{1}{\sqrt{6}}
~,~\text{and}\\
\left<\epsilon_A|\eta_C\right> & = e^{i\phi_{AC}}\frac{1}{\sqrt{6}} ~,\\
\end{align*}
\begin{align*}
\left<\epsilon_B|\eta_A\right> & = e^{i\phi_{BA}}\frac{1}{\sqrt{6}} ~,\\
\left<\epsilon_B|\eta_B\right> & = e^{i\phi_{BB}}\sqrt{\frac{2}{3}}
~,~\text{and}\\
\left<\epsilon_B|\eta_C\right> & = e^{i\phi_{BC}}\frac{1}{\sqrt{6}}
  ~,
\end{align*}
and:
\begin{align*}
\left<\epsilon_C|\eta_A\right> & = e^{i\phi_{CA}}\frac{1}{\sqrt{6}} ~,\\
\left<\epsilon_C|\eta_B\right> & = e^{i\phi_{CB}}\frac{1}{\sqrt{6}} ~,\\
\left<\epsilon_C|\eta_C\right> & = e^{i\phi_{CC}}\sqrt{\frac{2}{3}}
  ~.
\end{align*}
We list the available geometric symmetries that leave the final stationary state unchanged:
\begin{enumerate}
\setlength{\topsep}{0pt}
\setlength{\itemsep}{0pt}
\setlength{\parsep}{0pt}
\item Phase transformation on each state, $\left|\eta_x\right>\mapsto e^{i\phi_x}\left|\eta_x\right>$;
\item Phase transformation on each dual state, $\left|\epsilon_x\right>\mapsto
e^{i\phi_x}\left|\epsilon_x\right>$; and
\item Unitary transformation $\left|\eta_x\right> \mapsto U\left|\eta_x\right>$ and $\left<\epsilon_x\right| \mapsto \left<\epsilon_x\right|U^\dagger$.
\end{enumerate}
From these symmetries we can fix gauge in the following ways:
\begin{enumerate}
\setlength{\topsep}{0pt}
\setlength{\itemsep}{0pt}
\setlength{\parsep}{0pt}
\item Set $\left<0|\eta_x\right>$ to be real and positive for all $x$.
\item Set $\phi_{AA}=\phi_{BB}=\phi_{CC}=0$.
\item Set $\left<0|\eta_A\right>=0$ and set $\left<1|\eta_B\right>$ to be real and positive.
\end{enumerate}

These gauge fixings allow us to write:
\begin{align*}
\left|\eta_A\right> & = \left|0\right> ~, \\
\left|\eta_B\right> & = \alpha_B \left|0\right> +\beta_B\left|1\right> ~,
~\text{and}\\
\left|\eta_C\right> & = \alpha_C \left|0\right> +e^{i\theta}\beta_C\left|1\right>
  ~,
\end{align*}
for $\alpha_B,\alpha_C\geq 0$, $\beta_B=\sqrt{1-\alpha_B^2}$ and $\beta_C=\sqrt{1-\alpha_C^2}$ and a phase $\theta$.

That these states are embedded in a 2D Hilbert space means
there must exist some linear consistency conditions.
For some triple of numbers
$\mathbf{c}=(c_A, c_B, c_C)$ we can write:
\begin{align*}
c_A\left|\eta_A\right> +c_B\left|\eta_B\right> +c_C\left|\eta_C\right> = 0
  ~.
\end{align*}
Up to a constant, we use our parameters to choose:
\begin{align*}
(c_A, c_B, c_C) = \left(e^{i\theta}\alpha_B\frac{\beta_C}{\beta{B}}-\alpha_C,\,-e^{i\theta}\frac{\beta_C}{\beta{B}},\,1\right)
  ~.
\end{align*}
Consistency requires that this relationship between vectors is preserved
by the Kraus operator dynamic. 
Consider the matrix $\mathbf{A}=(A_{xy}) = \left(\left<\epsilon_x|\eta_y\right>\right)$. 
The vector $\mathbf{c}$ must be a null vector of $\mathbf{A}$; 
i.e. $\sum_y A_{xy}c_y=0$. This first requires that $A_{xy}$ be degenerate. 
One way to enforce this to check that the characteristic polynomial 
$\det(\mathbf{A}-\lambda \mathbf{I}_3)$ has an overall factor of $\lambda$. 
For simplicity, we compute the characteristic polynomial of $A\sqrt{6}$:
\begin{align*}
& \det(\sqrt{6}\mathbf{A}-\lambda \mathbf{I}_3) = \\
  & \quad \left(2-\lambda\right)^3 + \\
  & \quad
\left(e^{i(\phi_{AB}+\phi_{BC}+\phi_{CA})}+e^{i(\phi_{BA}+\phi_{CB}+\phi_{AC})}\right)
 - \\
& \quad (2-\lambda)\left(e^{i(\phi_{AB}+\phi_{BA})}+e^{i(\phi_{AC}+\phi_{CA})}+e^{i(\phi_{BC}+\phi_{CB})}\right)
  ~.
\end{align*}
To have an overall factor of $\lambda$, we need:
\begin{align*}
0 & = 8 +
  \left(e^{i(\phi_{AB}+\phi_{BC}+\phi_{CA})}+e^{i(\phi_{BA}+\phi_{CB}+\phi_{AC})}\right) \\
& \quad
-2\left(e^{i(\phi_{AB}+\phi_{BA})}+e^{i(\phi_{AC}+\phi_{CA})}+e^{i(\phi_{BC}+\phi_{CB})}\right)
  ~.
\end{align*}
Typically, there will be several ways to choose phases to cancel out
vectors, but in this case since the sum of the magnitudes of the complex terms
is 8, the only way to cancel is at the extreme point where
$\phi_{AB}=-\phi_{BA}=\phi_1$, $\phi_{BC}=-\phi_{CB}=\phi_2$, and
$\phi_{CA}=-\phi_{AC}=\phi_3$ and:
\begin{align*}
\phi_{1}+\phi_2+\phi_3=\pi
  ~. 
\end{align*}

To recapitulate the results so far, $\mathbf{A}$ has the form:
\begin{align*}
\mathbf{A} = \frac{1}{\sqrt{6}}
\left(\begin{array}{ccc}
2 & e^{i\phi_1} & -e^{i(\phi_1+\phi_2)}\\
e^{-i\phi_1} & 2 & e^{i\phi_2}\\
-e^{-i(\phi_1+\phi_2)} & e^{-i\phi_2} & 2
\end{array}\right)
  ~.
\end{align*}
We now need to enforce that $\sum_y A_{xy}c_y=0$. We have the three equations:
\begin{align*}
2c_A+e^{i\phi_1}c_B -e^{i(\phi_1+\phi_2)}c_C & = 0 ~, \\
2c_B+e^{-i\phi_1}c_A +e^{i\phi_2}c_C & = 0 ~, ~\text{and}\\
2c_C+e^{-i\phi_2}c_B -e^{-i(\phi_1+\phi_2)}c_A & = 0
  ~.
\end{align*}
It can be checked that these are solved by
\begin{align*}
c_A & = e^{i(\phi_1+\phi_2)}c_C
  ~\text{and} \\
c_B & = - e^{i\phi_2}c_C
  ~.
\end{align*}
Taking our formulation of the $\mathbf{c}$ vector, we immediately have
$\beta_B=\beta_C=\beta$ (implying $\alpha_B=\alpha_C=\alpha$), $\phi_2 =
\theta$, and:
\begin{align*}
e^{-i\phi_3} & = \alpha(1-e^{i\theta}) \\
  & =-2i\alpha\sin(\theta)e^{i\theta/2} \\
  & =\alpha\sin(\theta)e^{i(\theta-\pi)/2}
  ~.
\end{align*}
This means:
\begin{align*}
\alpha & = \frac{1}{2} \left|\csc\left(\frac{\theta}{2}\right)\right|
~\text{and}\\
\phi_3 & = \frac{-\theta+\mathrm{sgn}(\theta)\pi}{2}
  ~,
\end{align*}
where we take $-\pi\leq \theta\leq \pi$ and $\mathrm{sgn}(\theta)$ is the sign
of $\theta$. 

Note, however, that for $-\frac{\pi}{3}< \theta< \frac{\pi}{3}$,
we have $|\csc(\theta)|> 1$, so these values are unphysical. Thus, we see that all parameters in our possible states $\left|\eta_x\right>$,
as well as all the possible transition phases, are dependent on the single
parameter $\theta$.
To construct the dual basis, we start with the new forms of the states:
\begin{align*}
\left|\eta_A\right> & = \left|0\right>, \\
\left|\eta_B\right> & = \alpha\left|0\right> + \beta\left|1\right> ~,
~\text{and}\\
\left|\eta_C\right> & = \alpha\left|0\right> + e^{i\theta}\beta\left|1\right>
  ~.
\end{align*}
We note directly that we must have:
\begin{align*}
\left<\epsilon_A|0\right> & = \sqrt{\frac{2}{3}}, \\
\left<\epsilon_B|0\right> & = \frac{1}{\sqrt{6}}e^{-i\phi_1} ~, ~\text{and}\\
\left<\epsilon_C|0\right> & = \frac{1}{\sqrt{6}}e^{i\phi_3}
  ~,
\end{align*}
from how the dual states contract with $\left|\eta_A\right>$. These can be used
with the contractions with $\left|\eta_B\right>$ to get:
\begin{align*}
\left<\epsilon_A|1\right> & =
\frac{1}{\beta}\sqrt{\frac{2}{3}}\left(\frac{1}{2}e^{i\phi_1}-\alpha\right) ~,\\
\left<\epsilon_B|1\right> &=
\frac{1}{\beta}\sqrt{\frac{2}{3}}\left(1-\frac{1}{2}\alpha e^{-i\phi_1}\right) ~,
~\text{and}\\
\left<\epsilon_C|1\right> & = \frac{1}{2\beta}\sqrt{\frac{2}{3}}\left(e^{-i\phi_2}-\alpha e^{i\phi_3}\right)
  ~.
\end{align*}
It is quickly checked that these coefficients are consistent with the action on
on $\left|\eta_C\right>$ by making liberal use of $e^{-i\phi_3} =\alpha(1-e^{i\theta})$.

Recall that with the correct dual states, the Kraus operators
take the form:
\begin{align*}
K_A & = \left|\eta_A\right>\left<\epsilon_A\right| ~, \\
  K_B & = \left|\eta_B\right>\left<\epsilon_B\right| ~, ~\text{and}\\
  K_C & = \left|\eta_C\right>\left<\epsilon_C\right|
  ~.
\end{align*}
Completeness requires:
\begin{align*}
\left|\epsilon_A\right>\left<\epsilon_A\right| + \left|\epsilon_B\right>\left<\epsilon_B\right| + \left|\epsilon_C\right>\left<\epsilon_C\right| = I
  ~.
\end{align*}
Define the vectors $u_x = \left<\epsilon_x | 0\right>$ and $v_x =
\left<\epsilon_x | 1\right>$. One can check that the above relationship implies
$\sum_x u_x^\ast u_x = \sum_x v_x^\ast v_x = 1$ and $\sum_x u_x v^\ast_x = 0$.
However, for our model, it is straightforward 
(though a bit tedious) to check that:
\begin{align*}
\sum_x u_x^\ast u_x & = \frac{2}{3}+\frac{1}{6}+\frac{1}{6} = 1 ~\text{and}\\
\sum_x v_x^\ast v_x & = \frac{1}{\beta^2}\left(1+\alpha^2-\alpha\cos\phi_1\right)
  ~.
\end{align*}
Using the definitions of $\alpha$, $\beta$, and $\phi_1$, the second equation can be simplified to:
\begin{align*}
\sum_x v_x^\ast v_x = \frac{2+\csc^2\frac{\theta}{2}}{4-\csc^2\frac{\theta}{2}}
  ~.
\end{align*}
This is unity only when $\csc^2\frac{\theta}{2}=1$, which requires that $\theta
= \pi$. This is, indeed, the model $\mathbf{D}_3$ that we have already seen.
 
This establishes that the only two-dimensional pure-state quantum model which
reproduces the 3-state MBW process is the one with a nonminimal statistical
memory $S(\rho_\pi)$. This means there cannot exist a quantum representation of
the 3-state MBW process that majorizes all other representations of the same.
For, if it existed, it must be a two-dimensional model and also minimize
$S(\rho_\pi)$.

\end{document}